\documentclass[aps,twocolumn,showpacs,amsmath,amssymb,prc]{revtex4-2}
\usepackage{graphicx,here}
\usepackage{enumitem}
\usepackage{multirow}
\usepackage{tikz}
\usepackage{epstopdf}
\usepackage[percent]{overpic}
\usepackage{scrextend}
\usepackage{xcolor,xspace}
\usepackage[ulem=normalem]{changes}
\usepackage{hyperref}
\hypersetup{colorlinks=True,urlcolor=blue,linkcolor=blue,citecolor=blue,filecolor=black}

\colorlet{Changes@Color}{red}
\usepackage{dcolumn,color,footnote,bm,braket}
\usepackage{url,longtable,tabularx}
\usepackage{threeparttable}

\usepackage{fancybox}
\usetikzlibrary{matrix}

\newcommand\+{\dagger}
\newcommand\nb{N_{\rm B}}
\newcommand\pt{(p,t)}
\newcommand\tp{(t,p)}

\newcommand{\beo}{B(E3; 3^-_{1} \!\to\! 0^+_{1})}
\newcommand{\bqmin}{\beta_{20}^{\rm min}}
\newcommand{\bomin}{\beta_{30}^{\rm min}}
\newcommand{\blmin}{\beta_{\lambda0}^{\rm min}}

\newcommand{\bej}[4]{#1^+_{#2} \!\to\! #3^+_{#4}}
\newcommand{\sigpt}[2]{I^{\pt}(0^+_{#1} \!\to\! 0^+_{#2})}
\newcommand{\sigtp}[2]{I^{\tp}(0^+_{#1} \!\to\! 0^+_{#2})}
\newcommand{\mpts}[2]{M^{\pt}_{s}(0^+_{#1} \!\to\! 0^+_{#2})}
\newcommand{\mptf}[2]{M^{\pt}_{sff}(0^+_{#1} \!\to\! 0^+_{#2})}
\newcommand{\mtps}[2]{M^{\tp}_{s}(0^+_{#1} \!\to\! 0^+_{#2})}
\newcommand{\mtpf}[2]{M^{\tp}_{sff}(0^+_{#1} \!\to\! 0^+_{#2})}

\newcommand{\ptgd}[2]{^{#1}{\rm Gd}(p,t)^{#2}{\rm Gd}}
\newcommand{\tpgd}[2]{^{#1}{\rm Gd}(t,p)^{#2}{\rm Gd}}

\newcommand{\tpnd}[2]{^{#1}{\rm Nd}(t,p)^{#2}{\rm Nd}}

\newcommand{\orme}[1]{\braket{3^-_1\|\hat T^{(E3)}\|0^+_{#1}}}

\newcommand{\nf}[3]{\braket{\hat n_f}_{{#1}^{#2}_{#3}}}

\begin{document}

\title{Octupole correlation effects on two-neutron
transfer intensity in rare-earth nuclei}

\author{Kosuke Nomura}
\email{nomura@sci.hokudai.ac.jp}
\affiliation{Department of Physics, 
Hokkaido University, Sapporo 060-0810, Japan}
\affiliation{Nuclear Reaction Data Center, 
Hokkaido University, Sapporo 060-0810, Japan}

\date{\today}

\begin{abstract}
Impacts of octupole correlations
on the low-lying $0^+$
states and two-neutron transfer
intensities in rare-earth nuclei are
investigated in terms of the interacting
boson model that is based
on the nuclear density functional
theory.
The octupole degrees of freedom
are not only essential building blocks
to describe properties of negative-parity
states in the model,
but also influence low-spin positive-parity
states including excited $0^+$ states.
The calculation produces a
large number of low-energy $0^+$ states
that contain significant
amounts of octupole components,
indicating important roles played
by the octupole degrees freedom
in this mass region.
Octupole correlations
are shown to make sizable
contributions to the $(p,t)$ and $(t,p)$
transfer intensities
and, in particular, to reproduce
the discontinuous
changes of these quantities
near those nuclei
with $N\approx88$ or 90,
which are observed experimentally
as a signature of
the shape phase transition.
\end{abstract}

\maketitle

\section{Introduction}

Reflection asymmetric, octupole
shapes in atomic nuclei have
attracted considerable interest
in the field of nuclear physics
\cite{ahmad1993,butler1996,
butler2016,butler2020b,butler2024}.
Up to the present,
evidence for a static octupole
shape was suggested
experimentally in a few nuclides
in light actinide
(e.g.,
$^{220}$Rn \cite{gaffney2013},
$^{224}$Ra \cite{gaffney2013}),
rare-earth and lanthanide (e.g,
$^{144}$Ba \cite{bucher2016},
and $^{146}$Ba \cite{bucher2017})
regions.
These nuclei exhibit a low-energy
negative-parity band,
which forms an
alternating-parity band with the
positive-parity ground-state band,
connected by strong electric
dipole ($E1$) and octupole ($E3$)
transitions.
The octupole correlations
are expected to be enhanced
for those nuclei with neutron
$N$ and/or proton $Z$
numbers approximately
equal to 34, 56, 88, 134,
$\ldots$,
for which the coupling
of single-particle states
that differ by $\Delta j=\Delta l=3$
is possible.
While there are only a few instances
that are empirically suggested
to exhibit the static octupole
deformation, the octupole
shape degrees of freedom
are still expected to be
a relevant ingredient
to determine various low-energy
nuclear structure phenomena.

In the light actinide and
rare-earth regions,
in addition to the low-energy
negative-parity bands that
are of octupole character,
low-lying excited $0^+$ states
and the bands built on them
have been observed.
The low-energy
$0^+$ states are often attributed
to the intruder states
that are different in
intrinsic structure from
the ground state
\cite{heyde1983,wood1992,
heyde2011,garrett2022,leoni2024}.
In addition, the observed
second $0^+$ energy levels
in rare-earth nuclei are known to
become particularly low
in the vicinity of
$N\approx90$,
and this systematic is
recognized as a signature
of the quantum phase transition
(QPT) \cite{casten2009,cejnar2010,carr-book,fortunato2021}
in nuclear shapes from nearly
spherical to strongly
prolate deformed
configurations.
Another interpretation of
the low-lying $0^+$ levels
has been made in terms of the
pairing degrees of freedom
\cite{bes1966,bes1972,broglia1973,
prochniak1999,
brink-broglia,
srebrny2006,
prochniak2007,
xiang2020,nomura2020pv}.
The presence of the $0^+$ bands
has also been considered to
be a consequence of the
coupling of two octupole phonons
(see Ref.~\cite{butler2016}
and references therein).

Here the two-neutron transfer,
i.e., $\pt$ and $\tp$,
reactions are of special interest.
These nuclear reactions
have been frequently
used in experiments to investigate
pairing properties and
collective excitations in
low-energy nuclear structure
\cite{BM,brink-broglia},
and should serve as a key tool
to understand nature of the
$0^+$ states in nuclei.
In particular, in recent years
$\pt$ and $\tp$ experiments
have been extensively
carried out to identify some new
low-lying states in actinide
\cite{levon2013,spieker2013,spieker2018}
and rare-earth
\cite{zamfir2002,meyer2006-pt,
pascu2009,pascu2010,bucurescu2019,
levon2020-158Gd-pt}
nuclei.
These works suggested that
the appearance of
a number of excited $0^+$
states at low energy,
and the structure of the
$0^+$ excited bands could
be interpreted to be, to a large
extent, the double octupole
phonon coupling.
Generally, drawing a robust conclusion
about the nature of the $0^+$
states and low-lying $K^{\pi}=0^+$ bands,
however, remains an open question.
To address the problem,
systematic theoretical studies
for nuclear spectroscopy
are in order, that are
based on a microscopic
nuclear structure model.

The interacting boson model (IBM)
\cite{IBM} is among the
viable nuclear structure models
able to give quantitative
descriptions of low-lying
states and two-nucleon
transfer reactions simultaneously.
The IBM has been successfully
used for the descriptions of
low-energy collective spectra
and transition
properties in medium-heavy
and heavy nuclei.
The assumption of the IBM,
in its simplest version, is that
the collective nucleon pairs with
spin and parity $J=0^+$
and $2^+$ are approximated by
monopole $s$ and quadrupole $d$ bosons,
respectively \cite{OAIT,OAI,IBM}.
The application of the
IBM to two-nucleon transfers
was made initially
in Ref.~\cite{arima1977trf}.
It has been further applied
to study $\pt$ and $\tp$
two-neutron transfer
reactions in the context of the shape QPTs
\cite{arima1977trf,scholten1978,
IBM,fossion2007,zhang2017,
nomura2019trf,garciaramos2020}.
The $\pt$ and $\tp$ intensities
calculated within the model exhibit
a phase-transitional, discontinuous
change at particular nucleon
numbers, and therefore can be
identified as signatures
of the QPTs.
The IBM calculations for the
two-neutron transfer intensities
were also made to support
the interpretation of the
$0^+$ excited states to be
of double-octupole phonon states
\cite{levon2013,spieker2013,spieker2018,
zamfir2002,meyer2006-pt,
pascu2009,pascu2010,bucurescu2019,
levon2020-158Gd-pt},
in which model parameters
were phenomenologically
fitted to the observed low-energy
spectra.

The present study
is aimed to investigate
possible effects of octupole correlations
on the low-energy $0^+$ states
and on the relevant two-neutron
transfer reactions in
the rare-earth region
using an updated version of
the IBM.
In the present work,
the Hamiltonian of the IBM is
completely determined by
means of the
constrained self-consistent
mean-field (SCMF) calculations
\cite{RS,bender2003}
based on a given nuclear
energy density functional (EDF):
The self-consistent calculations
provide the potential energy surface
(PES) in terms of the
relevant shape degrees of freedom,
and it is mapped onto
the corresponding energy
surface in the boson system
\cite{nomura2008,nomura2010}.
This method, hereafter
denoted mapped IBM,
has been successfully used
in a number of nuclear structure
(see Ref.~\cite{nomura2025rev}
and references therein),
$\beta$-decay
\cite{nomura2020beta-1,nomura2020beta-2,
nomura2024beta},
and double-$\beta$-decay studies
\cite{nomura2022bb,nomura2025bb-1,
nomura2025bb-2}.
In particular, the mapped IBM
has been extensively used for the
spectroscopic calculations
for the low-energy
quadrupole and octupole
collective excitations
in a wide range of the
nuclear chart
\cite{nomura2023oct}.

In Ref.~\cite{nomura2019trf},
the mapped IBM was implemented
in the calculations of the
$\pt$ and $\tp$ intensities
in the Sm, Gd, and Dy nuclei
with the microscopic inputs
provided by the Skyrme \cite{Skyrme}
SkM* \cite{SkM} EDF.
A problem of the IBM mapping
revealed in Ref.~\cite{nomura2019trf}
was that the
calculated $\pt$ and $\tp$
transfer intensities displayed
only monotonous changes
with the neutron number,
which contradicted the empirical
data showing a discontinuous
change at a particular nucleon
number typical of the
shape phase transitions.
A possible reason for this
is that the IBM model space
considered in \cite{nomura2019trf}
constitutes $s$ and $d$
bosons only.
It would be thus of interest
to see influences of including
the octupole degrees of freedom
in the IBM mapping calculations
on the systematic of the
transfer intensities.

In the present
study, the low-energy collective
states, $\pt$ and $\tp$ transfer
reactions in the even-even
Nd, Sm, and Gd
isotopes with $N=84-94$
are specifically considered.
These nuclei undergo
a rapid nuclear structure change
from nearly spherical
to strongly deformed shapes
near $N=90$,
and have been studied
in the IBM as an ideal testing
ground for the model.
The octupole
correlations are expected to
play a role in low-lying states
in these nuclei,
as they are close to
the neutron number $N=88$.
In addition, given that the mapped
IBM was previously
applied to the Gd, Sm \cite{nomura2015},
and Nd \cite{nomura2021oct-ba}
nuclei and provided
reasonable descriptions of
the low-lying quadrupole and
octupole collective states,
the ingredients
and results of these calculations
can be exploited,
namely, the IBM Hamiltonian
parameters derived from the
PES mapping.
For the microscopic part,
the SCMF calculations were
performed in
Refs.~\cite{nomura2015,nomura2021oct-ba}
within the
Hartree-Fock-Bogoliubov (HFB)
method \cite{robledo2019} using the
Gogny \cite{Gogny}
interaction D1M \cite{D1M}.

The paper is organized as follows.
The theoretical framework
of the mapped IBM including
the formulation of the
two-nucleon transfer operators
is described in Sec.~\ref{sec:model}.
Section~\ref{sec:results}
provides results of the
mapped IBM calculations
for low-energy
spectroscopic properties,
including the excitation energies
and $B(E3)$ values,
$\pt$ and $\tp$
transfer intensities.
Section~\ref{sec:summary}
is devoted to a summary
of the principal results
and possible future studies.

\section{Theoretical procedure\label{sec:model}}

In the present version of the IBM
\cite{IBM,engel1985,engel1987},
in addition to $s$ and $d$ bosons,
which produce positive-parity
states, octupole $f$ bosons,
reflecting collective $J=3^-$
nucleon pairs, should be included
to calculate negative-parity states.
To keep the discussion
simple, the neutron and
proton bosons are not distinguished,
unlike the previous work
of Ref.~\cite{nomura2019trf}.
The $sdf$-IBM Hamiltonian
takes the form
\cite{nomura2015,nomura2021oct-ba};
\begin{align}
\label{eq:ham}
\hat H= 
\epsilon_d\hat n_{d} 
+ \epsilon_{f}\hat{n}_{f} 
+ \kappa_{2}\hat{Q}_{2}\cdot\hat{Q}_{2}
+ \kappa_{3}\hat{Q}_{3}\cdot\hat{Q}_{3}
+ \rho\hat{L}\cdot\hat{L} \; .
\end{align}
The first and second terms,
given respectively as
$\hat n_d=d^{\+}\cdot\tilde d$
and $\hat n_f=f^{\+}\cdot\tilde f$,
denote $d$- and $f$-boson number
operators.
The third, fourth, and fifth
terms are quadrupole-quadrupole,
octupole-octupole, and rotational
terms, respectively,
with the corresponding
operators being
\begin{subequations}
 \begin{align}
\label{eq:q2}
& \hat Q_{2}=s^{\dagger}\tilde d+d^{\dagger}\tilde s+\chi_{d}(d^{\dagger}\tilde
  d)^{(2)}+\chi_{f}(f^{\dagger}\tilde f)^{(2)} \\
\label{eq:q3}
&\hat{Q}_{3}=
s^{\dagger}\tilde{f}+f^{\dagger}\tilde{s}
+\chi_{3}(d^{\dagger}\tilde{f}
+f^{\dagger}\tilde{d})^{(3)}
\\
\label{eq:l}
& \hat{L}=
\sqrt{10}(d^{\dagger}\tilde{d})^{(1)}
+\sqrt{28}(f^\+\tilde{f})^{(1)} \; .
\end{align}
\end{subequations}
The Hamiltonian parameters,
$\epsilon_d$, $\epsilon_f$,
$\kappa_2$, $\kappa_3$, $\rho$,
$\chi_d$, $\chi_f$, and $\chi_3$,
are determined in such a way
\cite{nomura2015,nomura2021oct-ba}
that the energy surface of the
$sdf$-IBM, which is provided
as the expectation value of the
Hamiltonian $\hat H$ in the
coherent state \cite{ginocchio1980}
of $s$, $d$, and $f$ bosons,
is made to match the
Gogny-HFB PES in terms of the
axially symmetric
quadrupole $\beta_{20}$
and octupole $\beta_{30}$
deformations.
Further details of the $sdf$-IBM
mapping procedure and derived
values of the parameters are found
in Refs.~\cite{nomura2015,nomura2021oct-ba}.
To study the $f$-boson
effects on spectroscopic properties,
another set of calculations using
the $sd$-IBM mapping is
also carried out, with the
corresponding Hamiltonian being
of the same form
as that in \eqref{eq:ham},
but with those terms involving
$f$ bosons omitted.

Within the IBM,
the $\pt$ and $\tp$ transfers
are represented by the removal
and addition of one boson,
respectively, if the boson
is treated as a like-particle pair.
The present study will be
focused on
the transfer reactions from the
$0^+_1$ ground-state state
in the initial nucleus to
the excited $0^+$ states in the
final nucleus,
hence the angular momenta are
unchanged ($\Delta I=0$).
The bosonic operators for
the $\pt$ and $\tp$ transfers
are here assumed to be of the forms
\begin{eqnarray}
\label{eq:pt}
\hat P^{\pt}
=
c_{1}
\left[
\hat A(\Omega_{\nu},N_{\nu})
\tilde s
 + c_2
\hat n_{f}
\tilde s
\right]
\end{eqnarray}
and
\begin{eqnarray}
\label{eq:tp}
\hat P^{\tp}
=
c_{1}
\left[
s^{\+}
\hat A(\Omega_{\nu},N_{\nu})
 + c_2
s^{\+}
\hat n_{f}
\right] \; ,
\end{eqnarray}
respectively,
where $\hat P^{\tp}
=\left(\hat P^{\pt}\right)^{\+}$,
$c_1$ and $c_2$
are parameters.
The operator
$\hat A(\Omega_\nu,N_\nu)$
is defined as \cite{arima1977trf}
\begin{eqnarray}
\label{eq:fac}
 \hat A(\Omega_\nu,N_\nu)
=
\left(
\Omega_\nu-N_\nu-\frac{N_\nu}{N_{\rm B}}\hat n_{d}
\right)^{1/2}
\left(
\frac{N_{\nu}+1}{\nb+1}
\right)^{1/2}
\; ,
\end{eqnarray}
where $\Omega_\nu$ denotes
the degeneracy
of the neutron pairs in a given major
shell, and is equal to
$(126-82)/2=22$ for the nuclei
under investigation.
$\nb$ is the total number
of bosons, and is equal to
the number of pairs of
valence nucleons \cite{OAIT,OAI,IBM}.
$N_{\nu}$ stands for the
number of like-neutron
bosons, i.e.,
$1\leqslant N_{\nu} \leqslant 6$
for those even-even nuclei with
$N=84,\ldots,94$, respectively.
As is conventional \cite{IBM},
the operator $\hat n_{d}$
in Eq.~(\ref{eq:fac}) is
replaced with its expectation
value in the ground state
of the initial nucleus,
i.e.,
$\braket{0^+_{1,i}|\hat n_{d}|0^+_{1,i}}$.
The $\pt$ and $\tp$ transfer
strengths are given as
\begin{align}
\label{eq:sigpt}
& \sigpt i f
=|\braket{N-2,0^+_f| \hat P^{\pt} | N,0^+_i}|^2
\\
\label{eq:sigtp}
& \sigtp i f
=|\braket{N+2,0^+_f | \hat P^{\tp} | N,0^+_i}|^2
\; .
\end{align}
$\ket{N,0^+_{i}}$
and $\ket{N\mp2,0^+_{f}}$ stand for
the wave functions for the
$0^+$ states in an initial
and a final nuclei
with the neutron numbers $N$ and
$N\mp2$ in the $\pt$
and $\tp$ reactions, respectively.
These wave functions
are obtained by diagonalizing
the mapped $sdf$-IBM Hamiltonian
in the boson
$m$-scheme basis \cite{nomura2022octcm}.

The transfer operators of the forms
\eqref{eq:pt} and \eqref{eq:tp}
are adopted from the $\pt$ studies in
actinides \cite{levon2013,spieker2013}
and in rare-earth
($^{166}$Er \cite{bucurescu2019}
and $^{158}$Gd \cite{levon2020-158Gd-pt})
regions,
in which phenomenological
calculations using the IBM
that consists of the $s$, $d$, $f$,
and $p$ (dipole) bosons
were carried out.
A difference from the transfer operators
considered in the $spdf$-IBM
calculations
is only that the single-$p$ boson term
is omitted in the present
calculation.
These phenomenological $spdf$-IBM
calculations suggested that,
in addition to the leading order
term [the first term in Eq.~\eqref{eq:pt}
or Eq.~\eqref{eq:tp}],
the transfer operators should
include terms involving the
negative-parity bosons such
as the second term in
\eqref{eq:pt} or \eqref{eq:tp}
to investigate the relevance
of the octupole correlations.
Unlike most of these empirical
IBM studies for the two-nucleon
transfers,
$p$ bosons are not considered
in the present work.
The reason for this is not
only to avoid complexities,
but is also that the
microscopic origin of $p$ bosons
has not been well established
as it is related to
a spurious center-of-mass motion
\cite{otsuka1986}
and including these degrees
of freedom may not be
fully justified.

%
%
\begin{figure}[htb!]
\begin{center}
\includegraphics[width=.9\linewidth]
{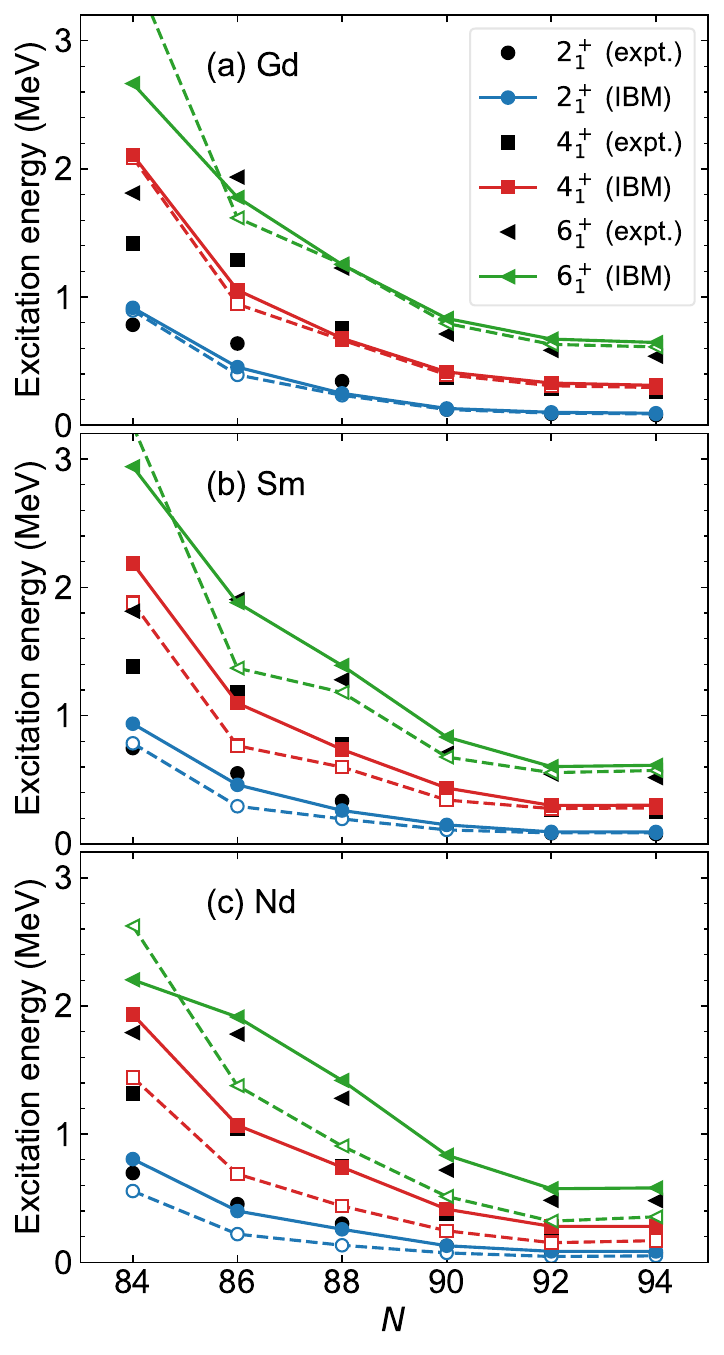}
\caption{
Calculated excitation energies for the
$2^+_1$, $4^+_1$, and $6^+_1$
states for the $^{148-158}$Gd,
$^{146-156}$Sm,
and $^{144-154}$Nd nuclei
within the mapped $sdf$-IBM
(solid symbols connected by sold lines)
and $sd$-IBM (open symbols connected
by dashed lines).
Experimental data are adopted
from NNDC database \cite{data}.
}
\label{fig:2+}
\end{center}
\end{figure}

%
%
\begin{figure}[htb!]
\begin{center}
\includegraphics[width=.9\linewidth]
{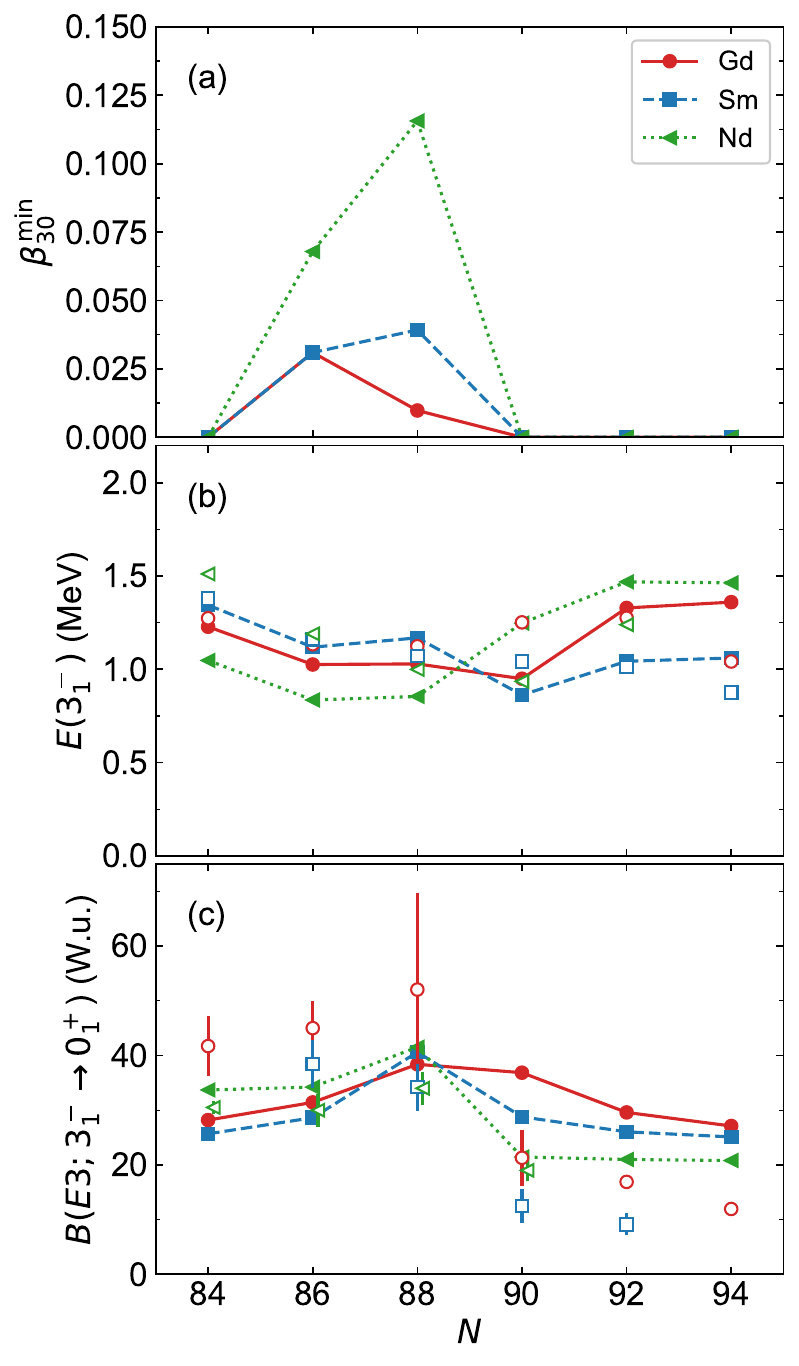}
\caption{
(a) Octupole deformation $\bomin$
corresponding to the energy minimum
in the Gogny-HFB
$(\beta_{20},\beta_{30})$-PES, 
(b) excitation energies $E(3^-_1)$,
and (c) $\beo$ values
for the Gd, Sm, and Nd isotopes
resulting from the mapped
$sdf$-IBM.
Experimental data for the $E(3^-_1)$
and $\beo$ values are taken
from Refs.~\cite{data,pascu2025prl},
and are represented by the
open symbols.
}
\label{fig:3-}
\end{center}
\end{figure}

%
%
\begin{figure*}[htb!]
\begin{center}
\includegraphics[width=.8\linewidth]
{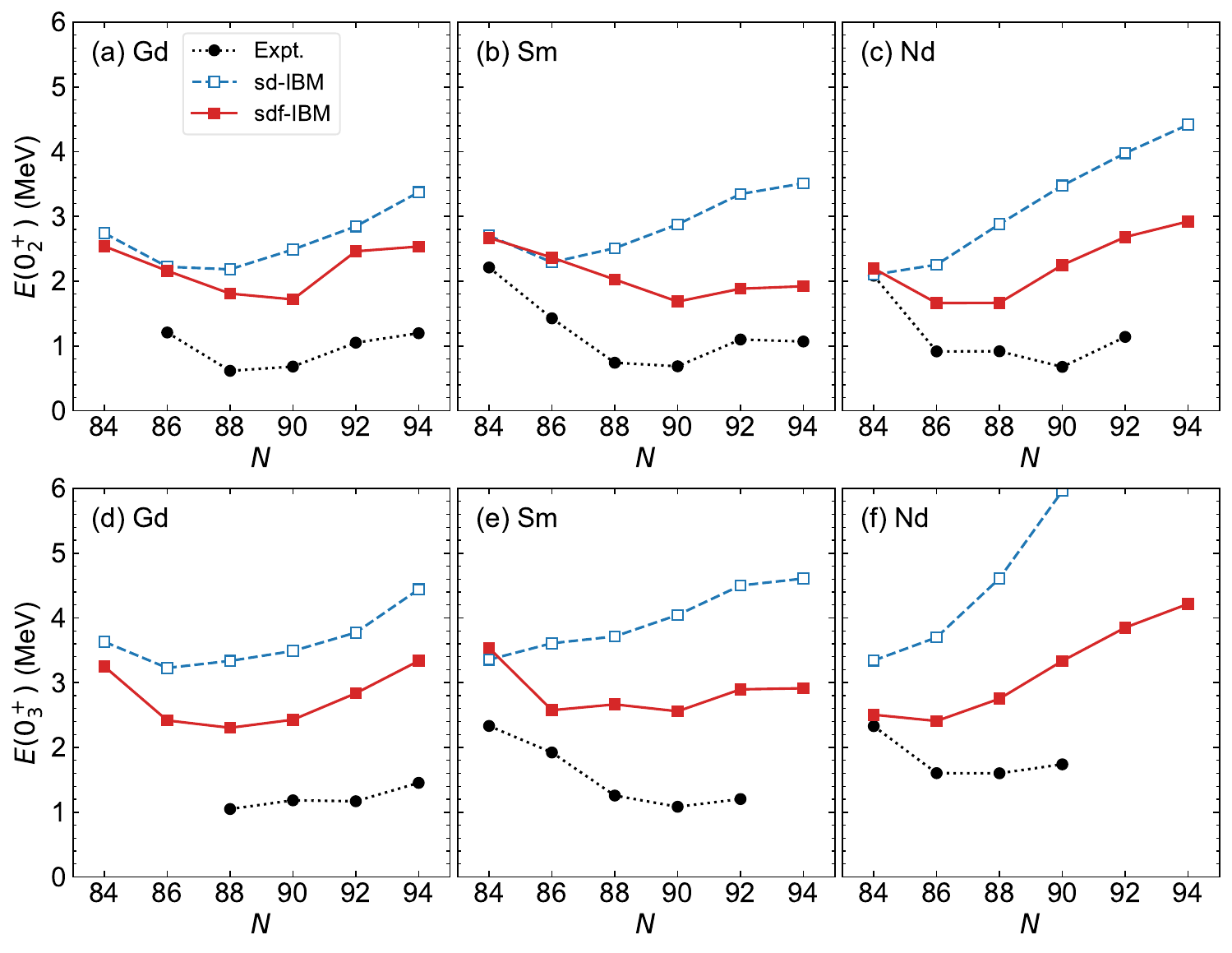}
\caption{
Calculated $0^+_2$ and $0^+_3$ energy
levels for the Gd, Sm, and Sm
isotopes within the mapped $sd$-IBM
and $sdf$-IBM in comparison with the
experimental data \cite{data}.
}
\label{fig:0+}
\end{center}
\end{figure*}

%
%
\begin{figure}[htb!]
\begin{center}
\includegraphics[width=.8\linewidth]
{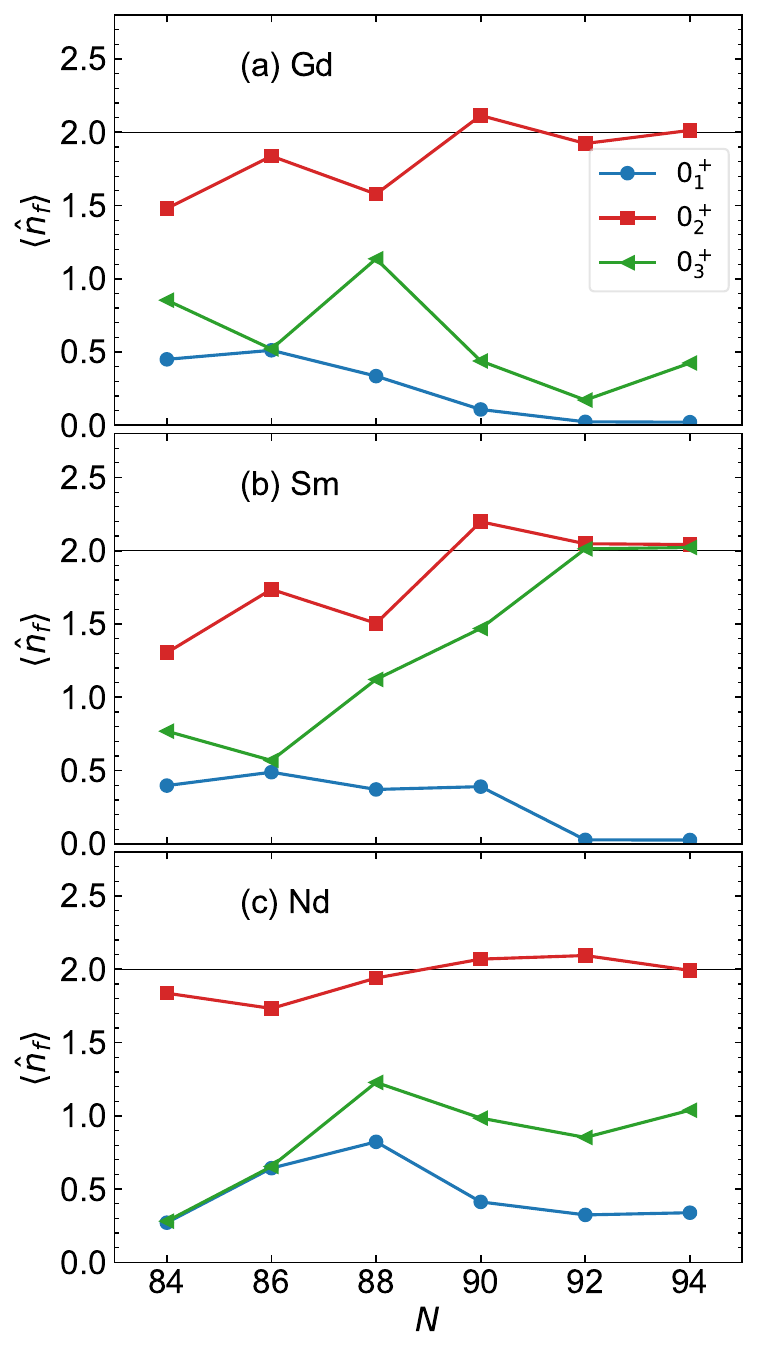}
\caption{
Expectation values of the
$f$-boson number operator $\braket{\hat n_f}$
in the $0^+_{1}$, $0^+_2$, and $0^+_3$
states.
}
\label{fig:nf-0+}
\end{center}
\end{figure}

%
%
\begin{figure}[htb!]
\begin{center}
\includegraphics[width=.8\linewidth]
{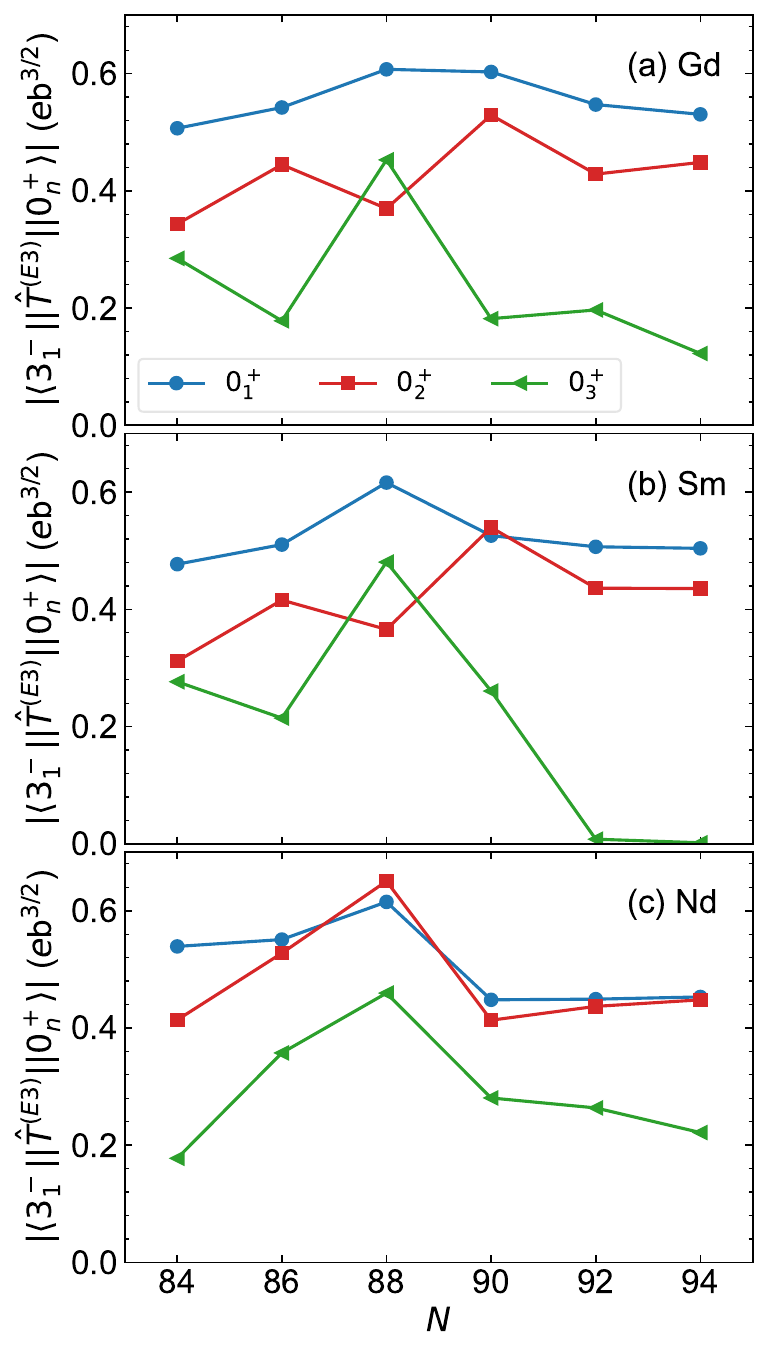}
\caption{
Absolute values of the
calculated reduced matrix elements
$|\orme n|$ ($n=1,2,3$) in $e$b$^{3/2}$ units.
}
\label{fig:e3rme}
\end{center}
\end{figure}

%
%
\begin{figure}[htb!]
\begin{center}
\includegraphics[width=\linewidth]
{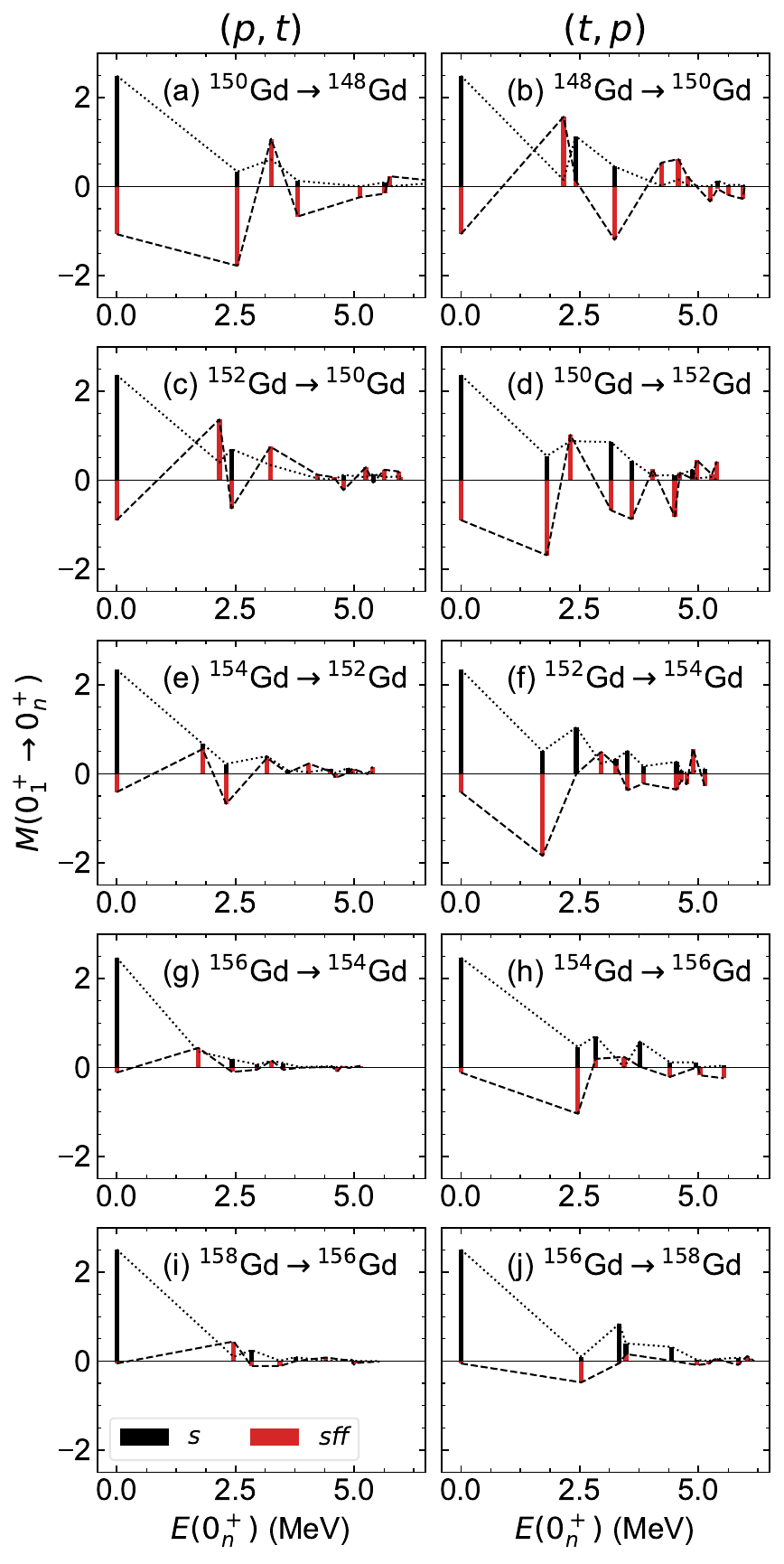}
\caption{
(Left) Matrix elements
$\mpts 1 n$ and $\mptf 1 n$
(denoted ``$s$'' and ``$sff$'',
and connected by dotted
and dashed lines,
respectively)
for the
$\ptgd {A} {A-2}$
transfer reactions
with $A=150-158$
calculated within the mapped $sdf$-IBM.
(Right) Matrix elements
$\mtps 1 n$ and $\mtpf 1 n$
for the $\tpgd {A} {A+2}$
reactions with $A=148-156$.
In the plot,
the sign of the matrix elements
$\mpts 1 n$ [$\mtps 1 n$]
are made to be positive, with
their relative sign to
$\mptf 1 n$ [$\mtpf 1 n$]
being unchanged.
}
\label{fig:me-gd}
\end{center}
\end{figure}

%
%
\begin{figure}[htb!]
\begin{center}
\includegraphics[width=\linewidth]
{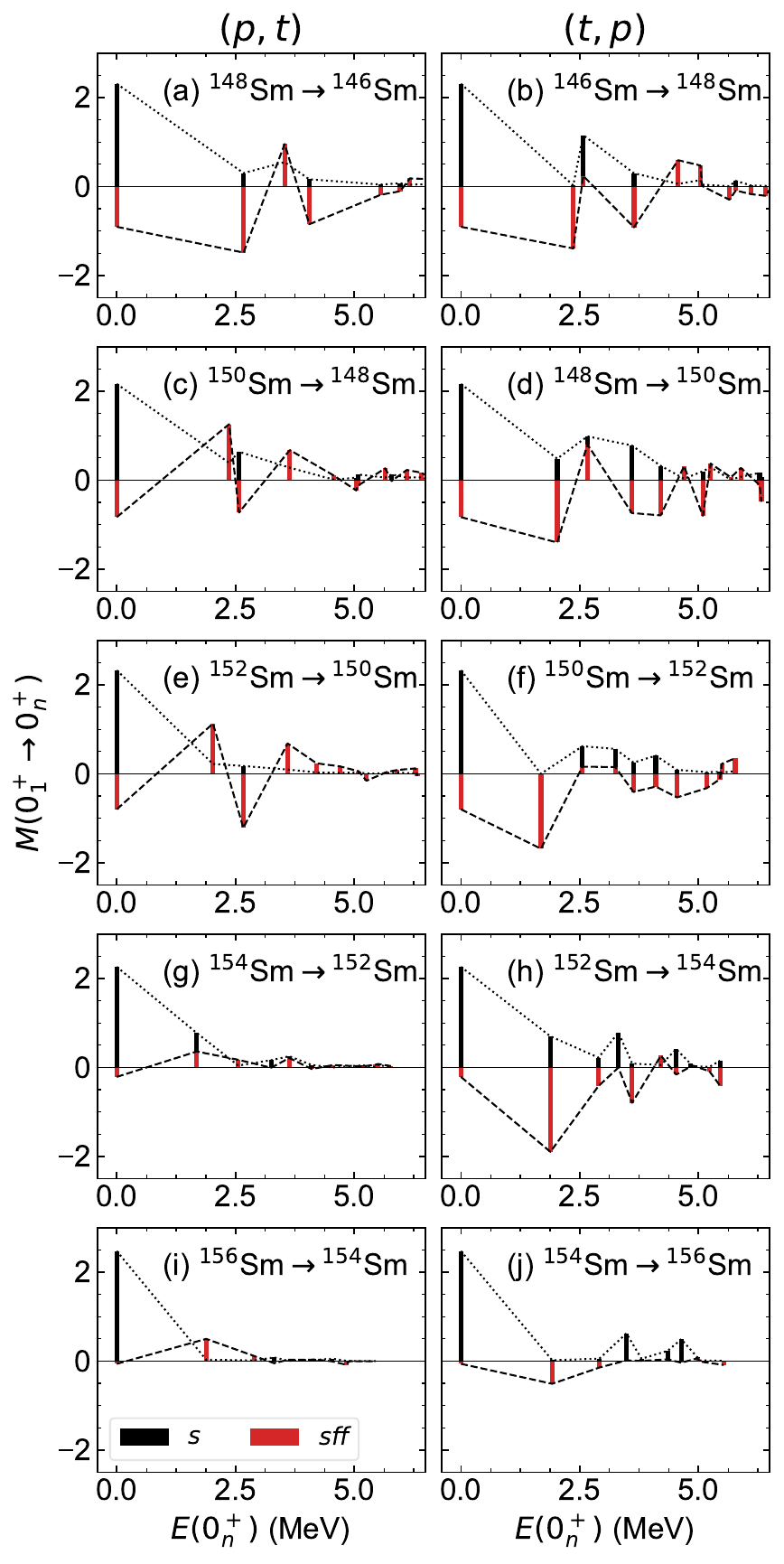}
\caption{
Same as the caption to Fig.~\ref{fig:me-gd},
but for the Sm isotopes.
}
\label{fig:me-sm}
\end{center}
\end{figure}

%
%
\begin{figure}[htb!]
\begin{center}
\includegraphics[width=\linewidth]
{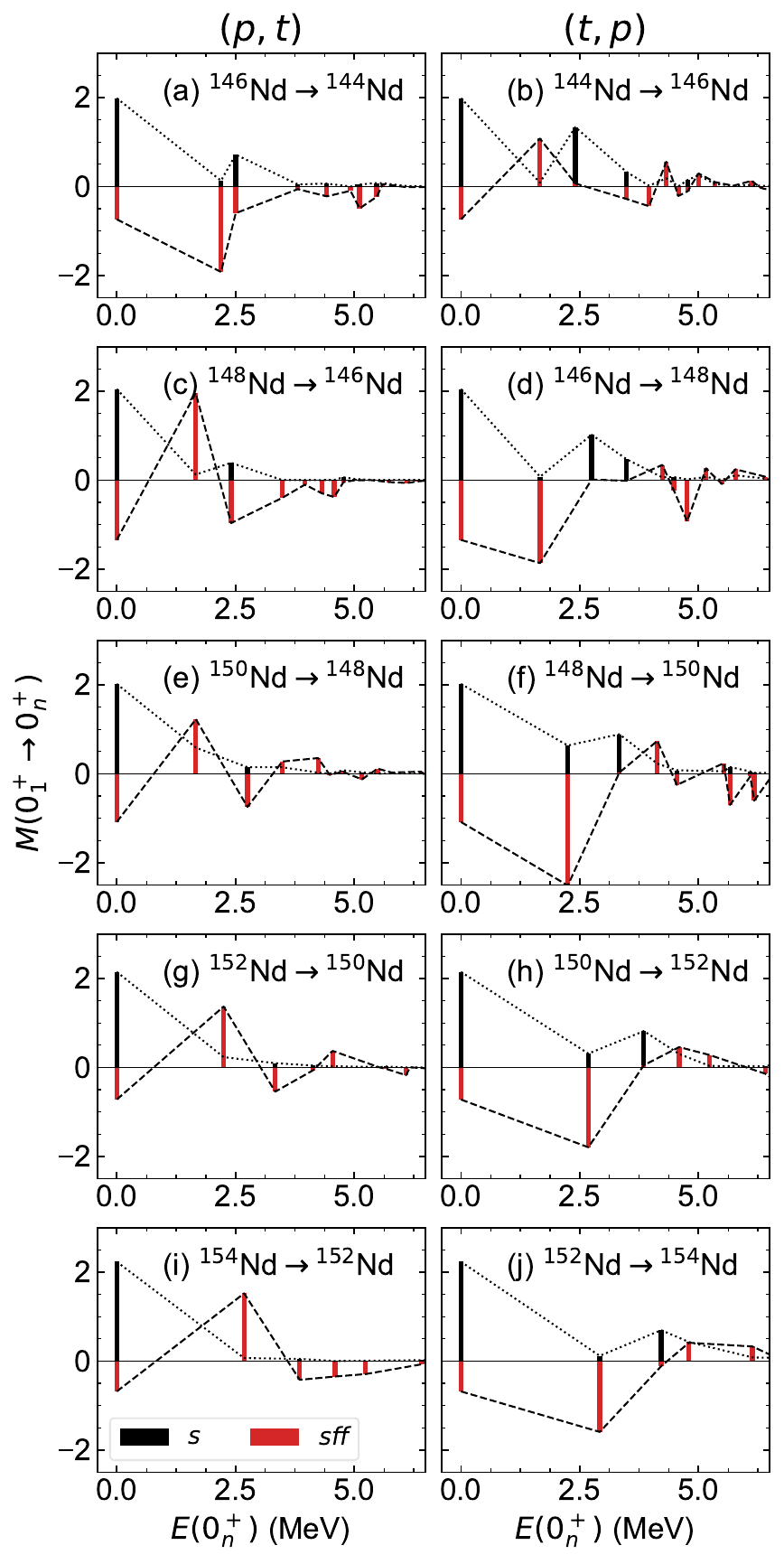}
\caption{
Same as the caption to Fig.~\ref{fig:me-gd},
but for the Nd isotopes.
}
\label{fig:me-nd}
\end{center}
\end{figure}

%
%
\begin{figure*}[htb!]
\begin{center}
\includegraphics[width=.8\linewidth]
{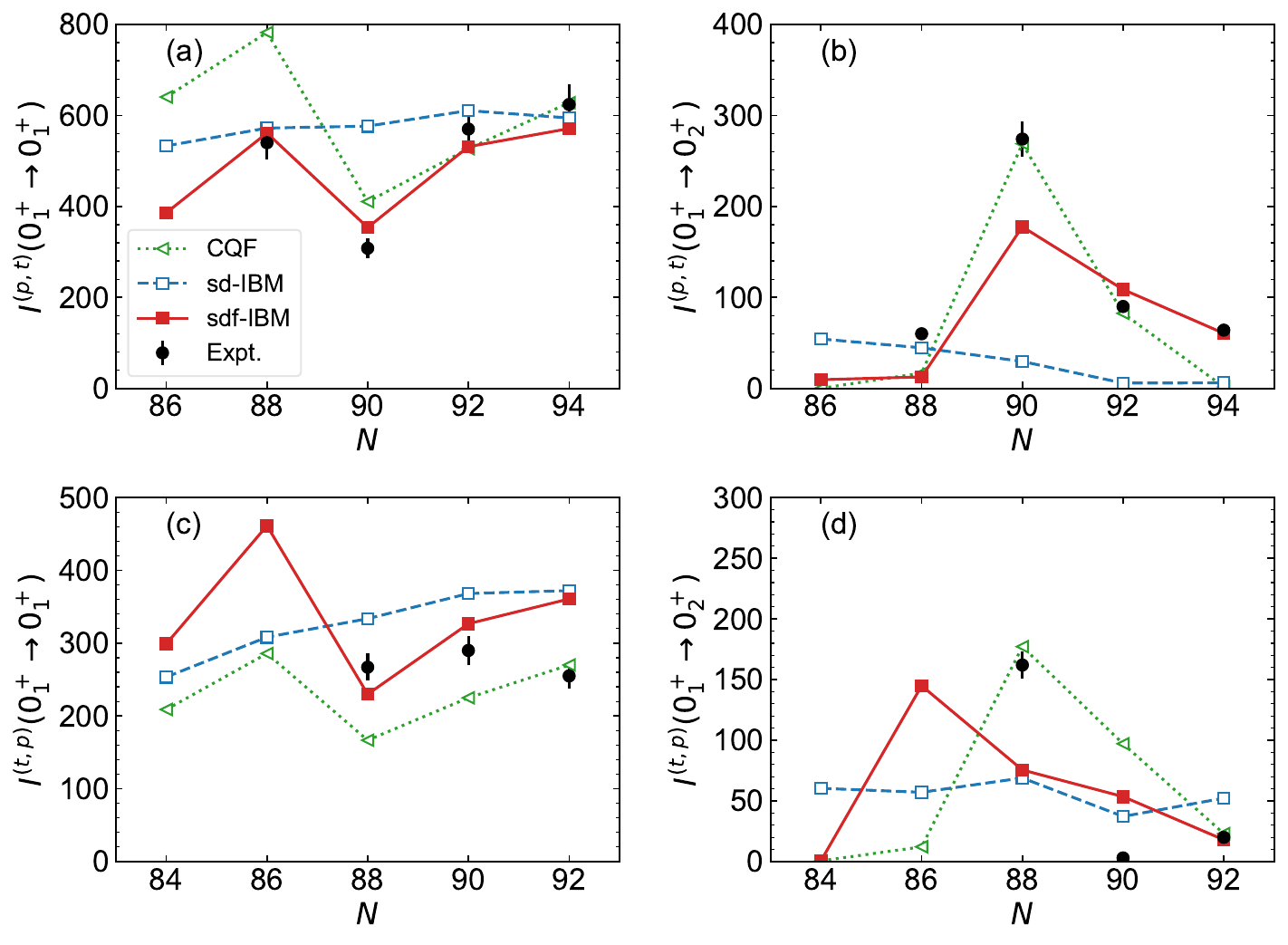}
\caption{
$\pt$ and $\tp$ transfer intensities
for the $\bej 0 1 0 1$
and $\bej 0 1 0 2$ transitions
in the Gd isotopes plotted
as functions of the neutron
number $N$ of the initial nucleus.
Calculations are made within the
mapped $sdf$-IBM and $sd$-IBM,
and phenomenological $sd$-IBM
using the CQF Hamiltonian.
Experimental data included are
differential cross
sections (in $\mu$b/sr units)
measured at the laboratory
angle $\theta_{\rm lab}=30^{\circ}$
for the $^{A}$Gd$\pt^{A-2}$Gd
\cite{fleming1973-Gd-pt},
and $^{A}$Gd$\tp^{A+2}$Gd
\cite{
SHAHABUDDIN1980-152Gd-tp,
LOVHOIDEN1989-154156Gd-tp}
reactions.
}
\label{fig:sigma-gd}
\end{center}
\end{figure*}

%
%
\begin{figure*}[htb!]
\begin{center}
\includegraphics[width=.8\linewidth]
{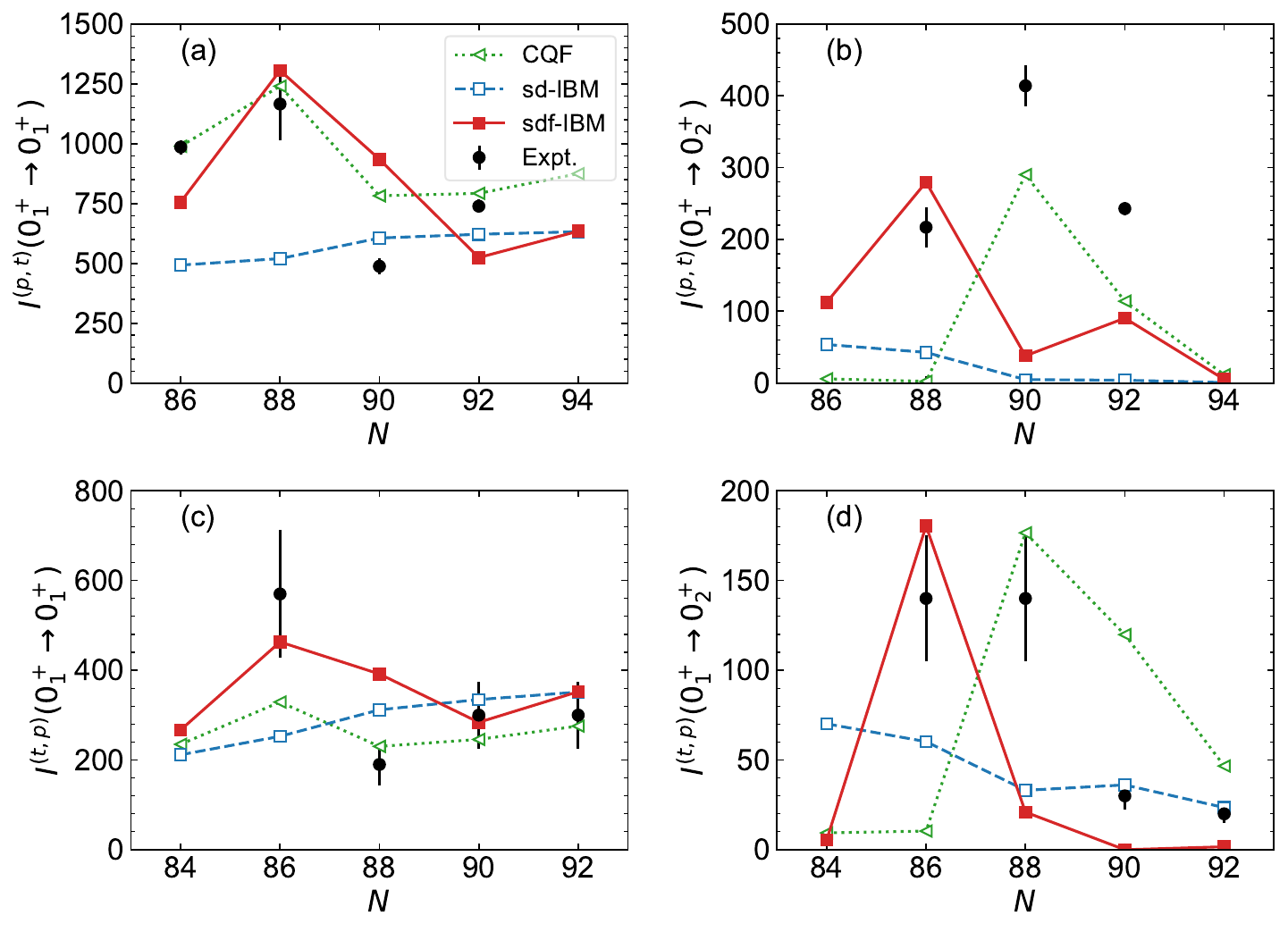}
\caption{
Same as the caption to
Fig.~\ref{fig:sigma-gd},
but for the Sm isotopic chain.
Experimental data for the
$\pt$ and $\tp$ transfers are
taken from
Refs.~\cite{DEBENHAM1972-Sm-pt,
BJERREGAARD1966-Sm-tp},
corresponding to the laboratory
angle $\theta_{\rm lab}=25^{\circ}$
and center-of-mass angle
$\theta_{\rm cm}=27.8^{\circ}$,
respectively.
}
\label{fig:sigma-sm}
\end{center}
\end{figure*}

%
%
\begin{figure*}[htb!]
\begin{center}
\includegraphics[width=.8\linewidth]
{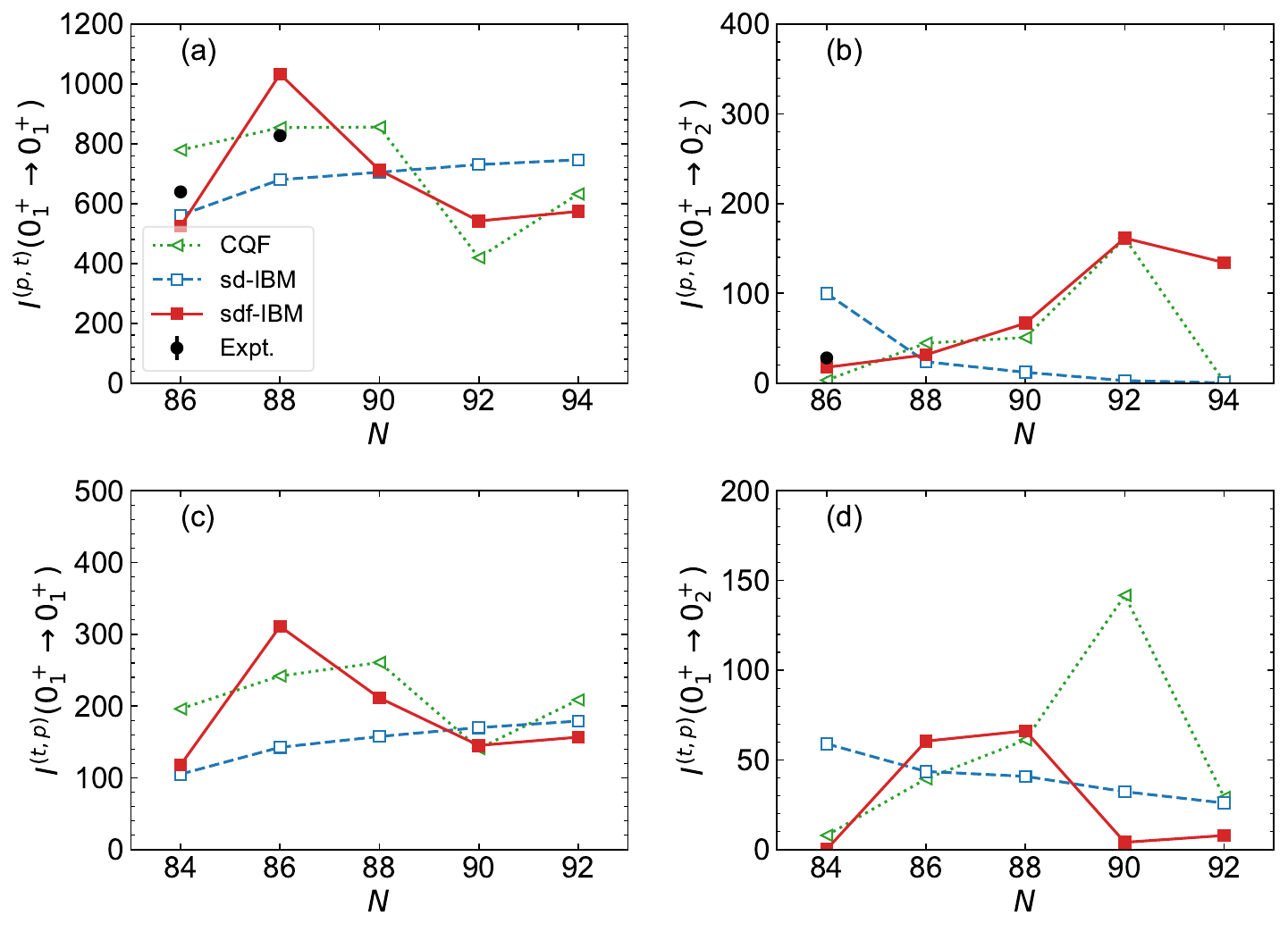}
\caption{
Same as the caption to
Fig.~\ref{fig:sigma-gd},
but for the Nd isotopic chain.
Experimental data for the $\pt$
transfers are adopted
from Ref.~\cite{PONOMAREV1996},
corresponding to the angle
$\theta_{\rm lab}=10^{\circ}$.
}
\label{fig:sigma-nd}
\end{center}
\end{figure*}

%
%
\begin{figure*}[htb!]
\begin{center}
\includegraphics[width=\linewidth]
{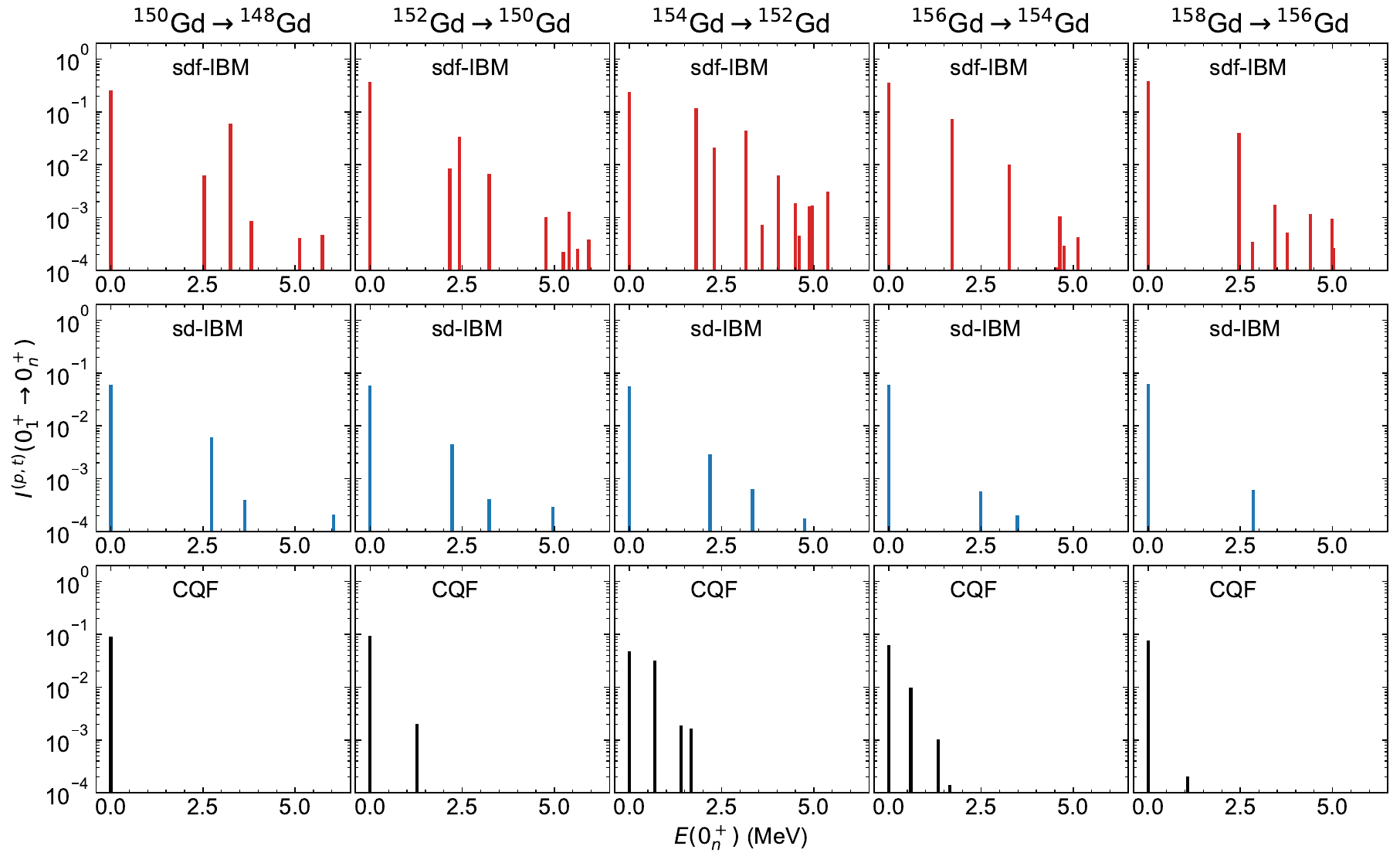}
\caption{
Predicted $\pt$
transfer intensities
$\sigpt 1 n$ from
the $sdf$-IBM, $sd$-IBM, and IBM-CQF
calculations for the Gd isotopes.
The adopted parameters for the
$\pt$ operators are
given in Table~\ref{tab:c1},
but the $c_1$ values
are replaced with a dimensionless
value $c_1=0.1$
for all the three IBM
calculations.
}
\label{fig:sdf-gd-pt}
\end{center}
\end{figure*}

%
%
\begin{figure*}[htb!]
\begin{center}
\includegraphics[width=\linewidth]
{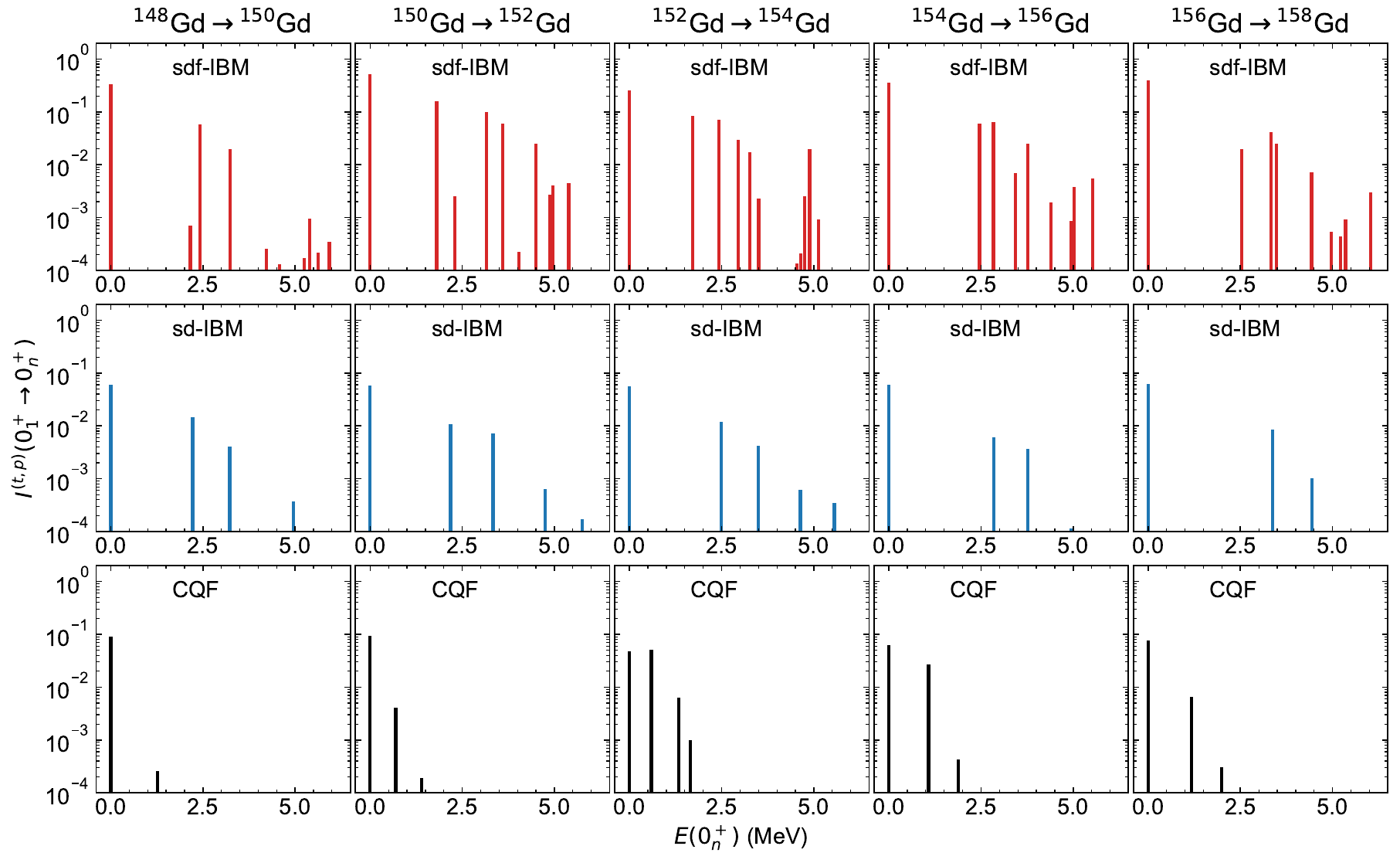}
\caption{
Same as the caption to
Fig.~\ref{fig:sdf-gd-pt}, but for
the $\tp$ transfer reactions.
}
\label{fig:sdf-gd-tp}
\end{center}
\end{figure*}

\section{Results\label{sec:results}}

\subsection{Low-energy spectroscopy\label{sec:level}}

Figure~\ref{fig:2+} depicts the
$2^+_1$, $4^+_1$, and $6^+_1$
energy levels computed within the
mapped $sdf$-IBM and $sd$-IBM.
These states constitute
the ground-state band, and
their decreases with $N$
indicate the increasing
quadrupole collectivity.
The Gogny-HFB
calculations carried out
in Refs.~\cite{nomura2015,nomura2021oct-ba}
suggested the increasing
quadrupole $\beta_{20}$
deformation with a
maximal value being
approximately 0.33 at $N=94$.
It is seen from Fig.~\ref{fig:2+}(a)
that there is no significant
difference between
the $sdf$-IBM and $sd$-IBM
results for the Gd isotopes with
$N\geqslant 88$,
hence the $f$-boson
effects on the ground-state-band
levels in Gd do not seem to be crucial.
The $sd$-IBM results for the
Sm and Nd nuclei,
shown in Figs.~\ref{fig:2+}(b)
and \ref{fig:2+}(c), generally
suggest lower energy levels
than the experimental ones \cite{data}.
The $sdf$-IBM
provides a higher and, overall,
more reasonable
description of excitation energies
for the yrast states than the $sd$-IBM.
The octupole correlation
effects on yrast spectra appear to be
particularly important in the Sm
and Nd isotopes.

Figure~\ref{fig:3-} shows
the calculated octupole
deformations, $\bomin$,
that correspond to the
global minima on the Gogny-HFB PESs,
and the excitation energies
of the $3^-_1$ state,
$E(3^-_1)$, and the
$\beo$ values,
resulting from the $sdf$-IBM.
Here the $E3$ operator $\hat T^{(E3)}$
is defined as
$\hat T^{(E3)}=e_3 \hat Q_3$,
with $e_3$ and $\hat Q_3$
being the effective boson charge and
octupole operator of \eqref{eq:q3},
respectively.
In Ref.~\cite{nomura2021oct-ba},
deformation-dependent $e_3$
charges of the form
\begin{eqnarray}
\label{eq:e3}
e_3=e_3^0(1+\overline{\bqmin}\,\overline{\bomin}) 
\end{eqnarray}
were considered
for Xe, Ba, Ce, and Nd isotopes.
Here $\overline{\blmin}$
denotes the bosonic
quadrupole ($\lambda=2$)
or octupole ($\lambda=3$)
deformation, and is assumed to be
proportional to the
fermion $\beta_{\lambda0}$
deformation,
$\overline{\blmin}
\equiv C_{\lambda}\blmin$,
with the constant
of proportionality $C_{\lambda}$.
$e^0_3$ is a scale factor
constant for an isotopic chain.
In the present calculations,
the $e_3$ charges in
the above formula \eqref{eq:e3}
are employed also for the
Gd and Sm isotopes,
and the scale factors
$e_3^0=0.105$ $e$b$^{3/2}$ for the
Gd and Sm isotopes,
and 0.12 $e$b$^{3/2}$
for the Nd isotopes are chosen
to reproduce to a good extent
the observed $\beo$ values.

One sees in Fig.~\ref{fig:3-}(a)
that nonzero
$\bomin$ values are suggested
for those nuclei with $N=86$ and 88,
with the maximal
octupole deformation being
$\bomin\approx 0.12$
for $^{148}$Nd.
For all isotopic chains,
the Gogny-HFB calculations
suggest that the octupole minimum
diminishes for those isotopes with
$N\geqslant 90$.
However, the PESs for the
heavier isotopes are also
$\beta_{30}$ soft
\cite{nomura2015,nomura2021oct-ba},
and thus suggest that
the dynamical octupole correlations
may still play a part for the
nuclei that are even away from
$N=88$.
The calculated $E(3^-_1)$
energies
exhibit a weak parabolic behavior
with a minimal value
at $N=86$ (Nd) or 90 (Gd and Sm),
near which the Gogny-HFB PESs exhibit
a $\beta_{30}\neq0$ minimum
or are $\beta_{30}$ soft.
The $sdf$-IBM gives a
systematic that is
more or less consistent with
experiment.
The lowering of the observed
$E(3^-_1)$ in Gd and Sm
from $N=92$ to 94
indicates that the octupole
correlations are still relevant
in these heavier isotopes.
The present $sdf$-IBM is,
however, unable to reproduce
this trend, mainly because the
Gogny-HFB PESs for
the heavier Gd and Sm isotopes
with $N\geqslant90$
do not exhibit a $\beta_{30}\neq0$
minimum.
Even though the Gogny-HFB
PESs show certain $\beta_{30}$
softness for these nuclei,
it does not seem to be fully
accounted for in the $sdf$-IBM.
Note that the $E(3^-_1)$ energies
for the $^{144,146,148}$Nd nuclei
are particularly low.
This is because, in these cases,
the Gogny-HFB PESs suggest
a too pronounced potential valley
along the $\beta_{30}$ deformation.
Corresponding to the predicted
$E(3^-_1)$ systematic,
the reduced transition rates,
$\beo$, exhibit an inverse parabola
with a maximal value
at $N\approx 88$,
confirming the relevance
of octupole collectivity.

Figure~\ref{fig:0+} compares the
excitation energies
$E(0^+_2)$ and $E(0^+_3)$
between the $sdf$-IBM
and $sd$-IBM.
The calculated $E(0^+_{2,3})$
in the $sd$-IBM are
considerably higher
than the experimental counterparts
by a factor of $2\sim 4$
particularly for the isotopes
with $N\geqslant 90$.
The overestimates of the
excited $0^+$ energy levels
are generally encountered
in the IBM mapping calculations,
and point to certain deficiencies
of the model, which
arise, e.g., from the restricted
boson model space
and/or the inputs provided by the
EDF self-consistent calculations.
The $sdf$-IBM provides
lower $0^+_{2,3}$ levels,
and reproduces qualitatively
the observed systematic more
reasonably than the $sd$-IBM.
It may thus appear
that the octupole degrees of freedom
could partly account for the
lowering of the excited
$0^+$ energy levels
in rare-earth nuclei.
However, it should be also
noted that the $sdf$-IBM
values for $E(0^+_{2,3})$
still overestimate
the experimental data,
and this implies that
some other correlations
need to be included in the IBM,
such as the configuration mixing,
which has been suggested
to be important in rare-earth region
(see, e.g.,
Refs.~\cite{garrett2009,heyde2011,tsunoda2023}).

Figure~\ref{fig:nf-0+} exhibits
expectation values of the $f$-boson
number operator $\hat n_f$
in the lowest three
$0^+$ states,
$\nf 0 + {n} \equiv\braket{0^+_n|\hat n_f|0^+_n}$,
obtained from
the mapped $sdf$-IBM.
For most of the nuclei,
$f$ bosons account
for small fractions
of the $0^+$ ground state
as the expectation values
are typically calculated to be
$\nf 0 + 1 \approx 0.5$.
For the $0^+_2$ states,
however, $\nf 0 + 2 \approx 2$
are predicted for most cases,
suggesting large octupole
contributions.
For the Sm and Gd nuclei,
a number of the calculated
higher-lying $0^+$
states in the $sdf$-IBM
were shown \cite{nomura2015}
to have approximately
two or even more $f$ bosons
in the corresponding wave functions,
$\nf 0 + n >2$.
This suggested that
octupole degrees of freedom play an
important role to explain
the existence of a large number
of the observed excited $0^+$ states
in this region.
Similar results
are obtained for the Nd isotopes.
The previous mapped $sdf$-IBM
calculations based on the
relativistic DD-PC1 EDF \cite{nomura2014}
and Gogny-D1M EDF \cite{nomura2020oct}
also suggested that more than
two $f$ bosons are contained
in the $0^+_2$ states
and the corresponding $K^{\pi}=0^+_2$
bands in light actinide regions.

Another signature of
octupole correlations
in a given excited $0^+$
state would be the $E3$ transition
matrix element $\orme {ex}$,
that is comparable in
magnitude to that
between the $0^+_1$ ground
state and $3^-_1$ state,
$\orme 1$.
Figure~\ref{fig:e3rme} depicts
the reduced $E3$ matrix elements,
$\braket{3^-_1\|\hat T^{(E3)}\|0^+_{n}}$,
for the first ($n=1$),
second ($n=2$), and third ($n=3$)
$0^+$ states predicted by
the mapped $sdf$-IBM.
The $E3$ matrix elements
for the ground states, $\orme 1$,
are generally largest
in each isotopic chain.
In many cases,
the $\orme 2$ values
are calculated to be of the
same order of magnitude
as, or even larger
(for $^{152}$Sm and $^{148}$Nd) than
the $\orme 1$ values,
confirming sizable
octupole contributions
to the $0^+_2$ states.
The $\orme 3$
matrix elements are, for most
of the nuclei, minor
in comparison with the
$\orme 1$ and $\orme 2$ values.
Specifically for $^{152}$Gd
and $^{150}$Sm,
$\orme 3$ are predicted to be larger
than the $\orme 2$
values.
Since the number of $f$
bosons in the $0^+_3$ and $3^-_1$
states for these nuclei are
predicted to be $\nf 0 + 3 \approx 1$
(Fig.~\ref{fig:nf-0+}),
and $\nf 3 - 1 \approx 1.5$
\cite{nomura2021oct-ba},
respectively,
such large $\orme 3$
matrix elements may have been obtained.
What is worth noting is
that nearly vanishing $\orme 3$
values are obtained
for $^{154}$Sm and $^{156}$Sm,
although the $0^+_3$
states in these nuclei are
similar to the $0^+_2$ states
in $f$-boson configuration
[cf. Fig.~\ref{fig:nf-0+}(b)].
Probably, it may
not be obvious to draw a firm
conclusion about the
$E3$ properties of the
$0^+_3$ states solely on the basis
of the $f$-boson content;
in fact, the $E3$ strengths $\orme 3$
for these deformed Sm nuclei
should also be sensitive to the
amounts of the $d$-boson
components in the
$3^-_1$ and $0^+_3$ states.

\subsection{Two-neutron transfer
intensities\label{sec:trf}}

By using the $sdf$-IBM wave functions
for each nucleus,
the two-neutron transfer
intensities defined in
Eqs.~\eqref{eq:sigpt} and \eqref{eq:sigtp}
are computed.
To see effects
of the octupole correlations,
the following matrix elements
are considered:
\begin{align}
\label{eq:mpts}
& {\mpts 1 n}
=\braket{N-2,0^+_n|\tilde s|N,0^+_1}\\
\label{eq:mptf}
& {\mptf 1 n}
=\braket{N-2,0^+_n|\hat n_f \tilde s|N,0^+_1}\\
\label{eq:mtps}
& {\mtps 1 n}
=\braket{N+2,0^+_n|s^{\+}|N,0^+_1}\\
\label{eq:mtpf}
& {\mtpf 1 n}
= \braket{N+2,0^+_n|s^{\+} \hat n_f|N,0^+_1}
\; .
\end{align}
The quantities in Eqs.~\eqref{eq:mpts}
and \eqref{eq:mptf}
[\eqref{eq:mtps}
and \eqref{eq:mtpf}]
correspond to
the first and second
terms of $\pt$ \eqref{eq:pt}
[$\tp$ \eqref{eq:tp}] transfer
operators, respectively,
up to the coefficients $c_1$
and $c_2$, and the factor
$\hat A_{\nu}(\Omega_\nu,N_\nu)$.
Figures~\ref{fig:me-gd}, \ref{fig:me-sm},
and \ref{fig:me-nd} depict
the calculated values of
these matrix elements
for the Gd, Sm, and Nd isotopes
plotted against the excitation
energies $E(0^+_n)$ ($n=1,2,\dots$)
in the final nuclei.

In all the three isotopic chains,
$\bej 0 1 0 1$
transfers are mostly dominated
by the $s$ boson terms,
that is, the matrix elements
$\mpts 1 1$ and $\mtps 1 1$
take particularly large
values.
However, the $f$-boson terms,
represented by $\mptf 1 1$ and $\mtpf 1 1$,
also make sizable contributions
in lighter Gd and Sm nuclei
with $N\leqslant88$
and many of the Nd nuclei.
Concerning the
$\bej 0 1 0 {2,3}$ transfers,
the $f$-boson contributions,
$\mptf 1 {2,3}$ and $\mtpf 1 {2,3}$,
are generally quite pronounced
with respect to the
single $s$-boson ones,
$\mpts 1 {2,3}$ and $\mtps 1 {2,3}$.
One can indeed see in
Figs.~\ref{fig:me-gd}, \ref{fig:me-sm},
and \ref{fig:me-nd}
that the corresponding
matrix elements
$\mptf 1 {2,3}$ and $\mtpf 1 {2,3}$
are much larger in magnitude
than $\mpts 1 {2,3}$ and $\mtps 1 {2,3}$
in majority of the cases.
Corresponding to the results
that large fractions of
octupole components enter
in the $0^+_2$ and $0^+_3$
states (cf. Fig.~\ref{fig:nf-0+}
and Ref.~\cite{nomura2015}),
the octupole degrees
of freedom play a relevant
role in the calculations of
two-neutron
transfer intensities
for the $\bej 0 1 0 {2,3}$ reactions.
As for those $0^+_n$ states higher
in energy than $0^+_3$,
the calculated matrix elements
\eqref{eq:mpts}--\eqref{eq:mtpf}
generally decrease in magnitude,
suggesting that the transfers to
higher-lying $0^+$ states
become less significant
with increasing excitation
energies.

\subsection{Signatures of shape phase transitions}

The $\pt$ and $\tp$ intensities
are now computed as signatures of
the shape phase transitions,
and are compared with
the experimental data.
As in literature \cite{zhang2017},
the experimental quantities to be
compared with the calculations
are those cross sections
measured at particular laboratory,
$\theta_{\rm lab}$, or
center-of-mass,
$\theta_{\rm cm}$, angles.
According to the results shown in
Figs.~\ref{fig:me-gd},
\ref{fig:me-sm}, and
\ref{fig:me-nd},
the $f$-boson effects
turn out to be pronounced
especially for the
$\bej 0 1 0 2$ $\pt$ and $\tp$
transfers for those
nuclei with $N\approx88$ and 90.
As shown in Fig.~\ref{fig:3-}(a),
the corresponding Gogny-HFB
PESs exhibit octupole $\beta_{30}$
minima.
To take these effects into
account, the coefficient
$c_2$ is here assumed to
depend on the $\beta_{30}$
deformation and vary with $N$:
\begin{eqnarray}
\label{eq:c2}
 c_2 = a + b[\beta_{30}^{\rm min}(N)
+\beta_{30}^{\rm min}(N\pm2)]/2
\; ,
\end{eqnarray}
with ``$-$'' for $\pt$ and
``$+$'' for $\tp$ transfers.
In the above formula,
$a$ and $b$ are parameters
constant for a given
isotopic chain and
for each of the $\pt$ and $\tp$
transfers,
and the second term
represents an average
$\bomin$ deformation of
the two neighboring isotopes.
The $a$ and $b$ values
are chosen so as to
reasonably reproduce the
overall behaviors of
experimental transfer intensities
with $N$.
The remaining parameter,
$c_1$
[\eqref{eq:pt} and \eqref{eq:tp}],
is an overall scale factor,
and is determined so that
the transfer intensities
should be
of the same order of magnitude
as the experimental values.
The adopted values of the parameters
$c_1$, $a$, and $b$
are summarized in Table~\ref{tab:c1},
and the resultant
$c_2$ values
are shown in Table~\ref{tab:c2}.
One can observe in Table~\ref{tab:c2}
that the $c_2$ values
vary appreciably at $N\approx 88$,
and take a large negative value
for the
$N=88\to86$ $\pt$ and
$N=86\to88$ $\tp$ transfers.

%
%
\begin{table}[hb!]
\caption{\label{tab:c1}
Adopted values of the
parameters for the transfer
operators in the $sdf$-IBM,
see Eqs.~\eqref{eq:sigpt}
and \eqref{eq:sigtp}.
}
\begin{center}
 \begin{ruledtabular}
\begin{tabular}{ccccccc}
& \multicolumn{2}{c}{Gd}
& \multicolumn{2}{c}{Sm}
& \multicolumn{2}{c}{Nd}
\\
\cline{2-3}
\cline{4-5}
\cline{6-7}
& $\pt$ & $\tp$
& $\pt$ & $\tp$
& $\pt$ & $\tp$
\\
\hline
$c_1$ &
3.9 & 2.5 & 3.9 & 2.8 & 4.7 & 2.3 \\
$a$ &
4 & 3.5 & 1 & 1 & 1.5 & 1 \\
$b$ &
$-200$ &$-250$ & $-150$ & $-100$ & $-25$ & $-30$\\
\end{tabular}
 \end{ruledtabular}
\end{center}
\end{table}

%
%
\begin{table}[hb!]
\caption{\label{tab:c2}
Values of the neutron-number-dependent
coefficients $c_2$ \eqref{eq:c2}
used for the $\pt$ and $\tp$
transfer operators
for the considered Gd,
Sm, and Nd nuclei.
}
\begin{center}
 \begin{ruledtabular}
\begin{tabular}{ccccccc}
& \multicolumn{2}{c}{Gd}
& \multicolumn{2}{c}{Sm}
& \multicolumn{2}{c}{Nd}
\\
\cline{2-3}
\cline{4-5}
\cline{6-7}
$N$
& $\pt$ & $\tp$
& $\pt$ & $\tp$
& $\pt$ & $\tp$
\\
\hline
$84$ & ${}$ & $-0.37$ & ${}$ & $-0.55$ & ${}$ & $-0.02$ \\ 
$86$ & $0.90$ & $-1.60$ & $-1.32$ & $-2.51$ & $0.65$ & $-1.75$ \\ 
$88$ & $-0.08$ & $2.28$ & $-4.26$ & $-0.96$ & $-0.79$ & $-0.73$ \\ 
$90$ & $3.02$ & $3.50$ & $-1.94$ & $1.00$ & $0.05$ & $1.00$ \\ 
$92$ & $4.00$ & $3.50$ & $1.00$ & $1.00$ & $1.50$ & $1.00$ \\ 
$94$ & $4.00$ & ${}$ & $1.00$ & ${}$ & $1.50$ & ${}$ \\
\end{tabular}
 \end{ruledtabular}
\end{center}
\end{table}

To study $f$-boson
effects quantitatively,
the $sd$-IBM mapping
calculations for the
$\sigpt 1 {1,2}$ and
$\sigtp 1 {1,2}$ values
are also performed.
For these calculations the
same values of the parameter
$c_1$ as those for the
$sdf$-IBM (cf. Table~\ref{tab:c1})
are adopted.

Furthermore,
the mapped $sdf$-IBM and $sd$-IBM
results are compared with those
obtained from purely
phenomenological $sd$-IBM
calculations in Ref.~\cite{zhang2017},
in which
the Hamiltonian in the so-called
consistent-Q formalism (CQF)
\cite{warner1983},
suitable for studying the shape QPTs,
was considered:
\begin{equation}
\label{eq:cqf}
 \hat H_\text{CQF}
=\epsilon_0
\left[
(1-\eta)\hat n_d -\frac{\eta}{4\nb}\hat Q^\chi\cdot\hat Q^\chi
\right]
\; .
\end{equation}
$\epsilon_0$ is
an overall scale factor usually
adjusted to reproduce
the experimental energy
levels in each nucleus, $\eta$ is
a control parameter,
and the operator
$\hat Q^{\chi}$ is the
quadrupole operator in
the $sd$-IBM with $\chi$
being the parameter
for the $(d^{\+}\tilde d)^{(2)}$
term.
The values of the $\epsilon_0$,
$\eta$, and $\chi$ parameters
are adopted from Tables~I, II,
and III of Ref.~\cite{zhang2017}.
As shown in Ref.~\cite{zhang2017},
the IBM-CQF fitting calculations
provide excellent phenomenological
descriptions of the experimental low-energy
spectra of the positive-parity
yrast states
and excited $0^+$ states
in rare-earth region.

Figures~\ref{fig:sigma-gd},
\ref{fig:sigma-sm}, and \ref{fig:sigma-nd}
show the calculated
$\sigpt 1 {1,2}$ and
$\sigtp 1 {1,2}$ intensities
in the Nd, Sm, and Gd isotopes.
Experimental data are
adopted from Refs.~
\cite{
SHAHABUDDIN1980-152Gd-tp,
LOVHOIDEN1989-154156Gd-tp,
DEBENHAM1972-Sm-pt,
BJERREGAARD1966-Sm-tp,
PONOMAREV1996}.
Some more details about the
data are given in the captions
to Figs.~\ref{fig:sigma-gd},
\ref{fig:sigma-sm},
and \ref{fig:sigma-nd}.
One can see from
Fig.~\ref{fig:sigma-gd}(a)
that the calculated $\sigpt 1 1$
value for Gd in the $sdf$-IBM
exhibits a discontinuous
change from $N=88-90$.
This systematic resembles
that obtained in the
IBM-CQF calculation, and
is considered to be
a signature of the
shape phase transition.
Algebraic relations in the IBM
suggest that, in the
large-boson-number limit ($\nb\to\infty$),
the $\bej 0 1 0 1$
two-neutron transfer intensities,
$|\braket{N,0^+_1|s|N+2,0^+_1}|^2$,
are proportional to
$\nb+1$ and $(\nb+1)(2\nb+3)/3(2\nb+1)$
in the U(5) (vibrational)
and SU(3) (rotational) limits,
respectively \cite{arima1977trf,IBM}.
In the IBM-CQF,
the phase-transitional
behavior of $\sigpt 1 1$
is reproduced solely with
the $s$ and $d$ boson degrees of
freedom,
using only the overall scale factor
$c_1$ in \eqref{eq:pt} and
\eqref{eq:tp} for the transfer operators.
In the EDF-mapped $sd$-IBM,
however, such an abrupt change
of the transfer intensities
cannot be produced,
provided that the transfer
operators are of the present forms
consisting only of the
term proportional to $s$ and
$s^{\+}$ operators.

In Fig.~\ref{fig:sigma-gd}(b)
the predicted $\sigpt 1 2$
value in the mapped $sdf$-IBM
exhibits a kink at $N=90$,
similarly to the IBM-CQF values.
Note that the selection rules
of the IBM with the $\nb\to\infty$
limit do not allow the
$\bej 0 1 0 2$ $\pt$
transitions in both the
U(5) and SU(3) limits.
However, as seen from
Fig.~\ref{fig:sigma-gd}(b),
all the present
IBM calculations
give finite $\sigpt 1 2$ values.
These deviations may
suggest that
the group theoretical argument
does not hold concerning
the $\sigpt 1 2$ intensity,
as the symmetries
are supposed to be broken
and certain degrees of
configuration mixing are
likely to take place
in realistic (numerical)
IBM calculations.
Experimental $\pt$ data for
the Gd nuclei and those for
the neighboring Sm
[Fig.~\ref{fig:sigma-sm}(b)]
also suggest large $\sigpt 1 2$
values.

The $\sigtp 1 1$ values,
displayed in Fig.~\ref{fig:sigma-gd}(c),
exhibit a quite similar $N$-dependence
to $\sigpt 1 1$ values.
The $\bej 0 1 0 1$
$\tp$ transfers are allowed by
the selection rules in the
IBM with $\nb\to\infty$ \cite{arima1977trf},
and the same symmetry limits
can be used to interpret
the change of $\sigpt 1 1$
from $N=86-88$ as the U(5)-SU(3)
phase transition.
In Fig.~\ref{fig:sigma-gd}(d),
the $\sigtp 1 2$ intensities
calculated in the IBM-CQF
show a similar systematic
to the $\sigpt 1 2$ one,
that is, the former is peaked
at $N=88$ and the latter at $N=90$.
The $sdf$-IBM also
gives a maximal
$\sigtp 1 2$ value at $N=86$,
exhibiting a certain similarity
in tendency to the
$\sigpt 1 2$ value.
The results of the three sets of
the IBM calculations that the
$\sigtp 1 2$ intensity
is enhanced for deformed or
transitional nuclei are more or less
consistent with the symmetry argument
in the IBM:
The matrix element
$|\braket{N+2,0^+_2|s^{\+}|N,0^+_1}|^2$
vanishes in the U(5) limit,
but is finite in the SU(3) limit
\cite{arima1977trf}.

The calculated
$\sigpt 1 1$ values in
the $sdf$-IBM for Sm isotopes,
shown in Fig.~\ref{fig:sigma-sm}(a),
exhibit a pattern
that signals the
nearly-spherical-to-deformed shape
phase transition,
as in the case of the Gd
isotopes.
This result is also overall
consistent with the
experimental data and
with the IBM-CQF fitting
calculation, apart from
the discrepancy in systematic
from $N=90-92$.
Figure~\ref{fig:sigma-sm}(b)
shows that the
$sdf$-IBM gives
the largest $\sigpt 1 2$
intensity at $N=88$,
whereas the IBM-CQF phenomenology
gives a vanishing $\sigpt 1 2$
intensity at $N=88$.
The experimental and IBM-CQF
$\sigpt 1 2$, however, become
maximal at $N=90$.
The discrepancy may be accounted
for by the fact that
the employed bosonic transfer
operators in \eqref{eq:pt} and
\eqref{eq:tp} are
of too simplified forms
to reproduce all details of
the empirical trends
of the transfer intensities,
or the chosen parameters for
the operators might not be
optimal.
In Fig.~\ref{fig:sigma-sm}(c),
the experimental data for the
$\sigtp 1 1$ intensities appear to show
a behavior similar to that of the
$\sigpt 1 1$ ones.
Both the mapped $sdf$-IBM
and IBM-CQF overall reproduce this
systematic.
As one can see in
Fig.~\ref{fig:sigma-sm}(d),
the mapped $sdf$-IBM predicts for
$\sigtp 1 2$ value to
be peaked at $N=86$, and
this value is within
error bar of
the experimental value
\cite{BJERREGAARD1966-Sm-tp}.

As one can see from
Fig.~\ref{fig:sigma-nd},
basically similar observations
can apply to the Nd isotopic chain.
The $\sigpt 1 {1,2}$
data are scarce and only
the relative intensities
of $\sigtp 1 {2}$
to $\sigtp 1 {1}$
are known experimentally
for the
$\tpnd {144} {146}$,
$\tpnd {146} {148}$,
$\tpnd {148} {150}$, and
$\tpnd {150} {152}$
transfer reactions
\cite{CHAPMAN1972-Nd-tp}.
The $a$ and $b$
parameters \eqref{eq:c2}
for the Nd isotopes
are thus determined
so as to be similar in tendency
to the $\sigpt 1 {1,2}$
values computed by the
IBM-CQF.
For the IBM-CQF values of
$\sigpt 1 {1,2}$ shown in
Figs.~\ref{fig:sigma-nd}(a)
and \ref{fig:sigma-nd}(b)
the $c_1$ value of 5.48 considered
in Ref.~\cite{zhang2017}
is here modified as $c_1=4.0$.
In addition, to compute the
$\sigtp 1 {1,2}$ intensities,
plotted in
Figs.~\ref{fig:sigma-nd}(c)
and \ref{fig:sigma-nd}(d),
the same $c_1$ parameter as
that for the Gd isotopes,
$c_1=2.24$ \cite{zhang2017},
is employed.
The measured $\sigtp 1 2/\sigtp 1 1$
ratios for the four reactions
are $0.09$, $0.15$, $1.28$,
and $0.72$, respectively,
obtained with the center-of-mass
angles $\theta_{\rm cm}=20.2^{\circ}$,
$27.8^{\circ}$, $27.8^{\circ}$,
and $27.8^{\circ}$.
These data suggest the
largest $\sigtp 1 2/\sigtp 1 1$
ratio for the
$\tpnd {148} {150}$ reaction,
and this systematic is
qualitatively
reproduced in the present
$sdf$-IBM calculation,
yielding the ratios
0.002, 0.19, 0.31, 0.03,
respectively.

\subsection{Transfer intensity distributions}

In Figs.~\ref{fig:sdf-gd-pt}
and \ref{fig:sdf-gd-tp},
calculated $\sigpt 1 n$
and $\sigtp 1 n$ transfer
intensities within the mapped
$sdf$-IBM are compared with
the corresponding quantities
obtained from the
mapped $sd$-IBM and IBM-CQF.
For the $sdf$-IBM, the $a$
and $b$ parameters given in
Table~\ref{tab:c1} are used.
The overall scale factor $c_1$
for all the three sets of the
IBM calculations is assumed
to be $c_1=0.1$,
in order to facilitate the
comparisons.
It is also noted that
only the calculated results
for Gd isotopes are discussed,
since the corresponding results
for the Sm and Nd isotopes
turn out be similar.

In Fig.~\ref{fig:sdf-gd-pt},
the $\sigpt 1 1$
values obtained from the
$sdf$-IBM
for the $\pt$ transfers
of the Gd isotopes are
all larger than those in
the $sd$-IBM and IBM-CQF
by an order of magnitude.
The $\sigpt 1 {2,3}$ values
for the transitional
nuclei near $N=88$ ($^{152}$Gd)
and 90 ($^{154}$Gd)
are predicted to be
larger in the $sdf$-IBM
than in the $sd$-IBM
by orders of magnitude.
Also a large number of excited
$0^+$ are predicted in the $sdf$-IBM
within the considered
energy range of up to 6.5 MeV.
In both the $sdf$-IBM and
$sd$-IBM, the $\pt$ intensities
are mostly accounted for by the
$\bej 0 1 0 1$ transfers
and, especially in the
$sd$-IBM, the contributions from
the higher-lying
$0^+$ states are
considerably minor.
In the IBM-CQF calculations,
the excited $0^+$ states are
generally predicted to be
much lower in energy than
those in the $sdf$-IBM
and $sd$-IBM mapping results.
As pointed out earlier,
the overestimates of the
$0^+_2$ and $0^+_3$ energies
in the mapped IBM calculations
imply certain missing correlations,
and necessitate further refinements
of the methodology.
Similarly to the two sets
of the mapped IBM calculations,
the IBM-CQF suggests that the
$\bej 0 1 0 1$ $\pt$ transfers
are dominant,
and the $\pt$ transfers to
higher-lying $0^+$ states
are essentially negligibly small.

Similar observations seem to
hold for the systematic of the
$\sigtp 1 n$ intensities.
In Fig.~\ref{fig:sdf-gd-tp}
there are also significant differences
in the $\sigtp 1 1$ values
between the $sdf$ and $sd$
boson models.
The $\tp$ transfer
to the $0^+_2$ state
in the $^{150}$Gd nucleus
is particularly strong within the
$sdf$-IBM, which is of the
same order of magnitude
as the $\sigtp 1 1$ transfer
intensity.
In the $sdf$-IBM, 
large $\tp$ strengths
to the higher-lying
$0^+$ states are obtained for
the $\tpgd {150} {152}$ and
$\tpgd {152} {154}$
transfer reactions.

\section{Conclusions\label{sec:summary}}

Influences of the octupole degrees
of freedom on the
two-neutron transfer reactions
and properties of excited $0^+$ states
in rare-earth nuclei
have been investigated
in terms of the $sdf$-IBM that is
based on the nuclear EDF.
In this framework,
the strength parameters
for the $sdf$-IBM Hamiltonian
were obtained by mapping the
axially symmetric quadrupole-octupole
energy surface calculated
by the constrained Gogny-EDF HFB
method onto
the expectation value of
the boson Hamiltonian
in the intrinsic state.
Resulting $sdf$-IBM
wave functions for the
initial- and final-state nuclei
were used to
calculate intensities of the
$\pt$ and $\tp$ transfers
between $0^+$ states.

While $f$ bosons are
essential building
blocks to compute negative-parity states,
these degrees of freedom
have been shown to play a role
in low-lying
positive-parity states,
including the $0^+_{2,3}$
energy levels.
The mapped $sdf$-IBM
calculations suggested that
the $0^+_2$ states in
almost all the considered
rare-earth nuclei
are accounted for by the presence
of $f$-boson components,
with the expectation
values typically
$\nf 0 + 2 \approx2$,
and that even the $0^+_1$
ground states in these nuclei
contain certain amounts
of the $f$-boson components.
As compared with the
$sd$-IBM, a large number
of the excited $0^+$ states
are produced in the $sdf$-IBM,
many of which were shown
to be characterized by
more than two $f$-boson content.
However, it should be also
noted that the calculated $0^+_{2,3}$
levels in the $sdf$-IBM
turned out to be
still much higher than
the experimental ones,
and some other degrees of
freedom, such as the
configuration mixing of
normal and intruder states,
and the dynamical pairing
deformations would need to
be taken into account.

For the $\bej 0 1 0 1$
transfer reactions,
the calculated matrix elements
of the leading-order
term in the transfer operator
proportional to $s(^{\+})$
have been shown to
be considerably larger than those
of the second term of the form
$\hat n_{f}s(^{\+})$.
However, the $\bej 0 1 0 {2,3}$
$\pt$ and $\tp$ matrix elements
of the $\hat n_{f}s(^{\+})$ term,
particularly for the
nuclei with $N=88$ or 90,
have been shown to be
of the same order of magnitude as
or even larger than
those of the single $s$-boson
[$s(^{\+})$] term.
The resultant
transfer intensities,
$\sigpt 1 {1,2}$ and
$\sigtp 1 {1,2}$, were
compared with the experimental
data and with those values obtained
from the mapped $sd$-IBM
and phenomenological
$sd$-IBM calculations.
The behavior of the $\pt$ and
$\tp$ transfer intensities are
considered to be empirical signatures
of the shape QPT from the
nearly spherical to strongly deformed
states in rare-earth region.
By assuming nucleon-number- and
deformation-dependent
parameters for the transfer operators,
the mapped $sdf$-IBM reproduced
the decreasing patterns of the
$\sigpt 1 {1,2}$ and
$\sigtp 1 {1,2}$ intensities
near $N=88$ or 90.
These results are consistent
with the experimental data
and with the IBM-CQF fitting
calculations.
In the EDF-mapped IBM framework,
the consideration of only the $s$ and $d$
boson degrees of freedom
turned out to be insufficient
to produce a discrete change
of the transfer intensities,
and the inclusion of $f$ bosons
appears to efficiently account
for such systematic.

The present work suggests the
relevance of octupole correlations
to predicting two-neutron
transfers, to interpretation
of the excited $0^+$ states
in terms of the multi-octupole-phonon
configurations,
and to the shape QPTs starting from
the microscopic EDF calculations.
A possible future work
is the extension of the
analysis to the $\pt$ and
$\tp$ studies in the
light actinide region, for which
updated experimental results
have been recently reported.
Also in this region,
a permanent octupole
deformation was empirically
suggested in a few nuclei,
and a similar type
of the nearly-spherical-to-axially-deformed
shape phase transitions
to those in the rare-earth region
are observed.
It would also be of interest
to assess and improve
the accuracy of the
theoretical method,
since, for instance,
the transfer operator
considered in this work is
of a simplified form,
and higher-order terms
may quantitatively change the
results and
influence conclusions.
In particular, given that the
IBM-CQF, involving only $s$ and $d$
bosons, can reproduce the
phase-transitional behaviors
of the two-nucleon transfer
intensities,
the mapped $sd$-IBM that includes
such high-order contributions
in the transfer operators
may be able to reproduce
the systematic of
the transfer intensities
without including $f$ bosons.
It would be an interesting
future study to investigate further
the roles played by the $f$
boson degrees of freedom
in the nuclear phenomena related
to the excited $0^+$ states
in deformed heavy nuclei
within the mapped IBM
In addition, given the important
roles played by the
configuration mixing of
intrinsic shapes in the
rare-earth region,
an extension of the IBM mapping
procedure to incorporate
mixing in the formalism
will be another interesting
future study.

\acknowledgements
This work has been supported by JSPS
KAKENHI Grant No. JP25K07293.

\bibliography{refs_v2}

\end{document}